\documentclass[graybox]{svmult}

\usepackage{mathptmx}       
\usepackage{helvet}         
\usepackage{courier}        
\usepackage{type1cm}        
\usepackage{makeidx}         
\usepackage{graphicx}        
\usepackage{multicol}        
\usepackage[bottom]{footmisc}

\usepackage{color}
\definecolor{keywordcolor}{rgb}{0,0,0}
\providecommand{\keyword}[1]{\index{#1}{\color{keywordcolor}#1}}

\makeindex             

\begin{document}

\title*{Narrowband Biphotons: Generation, Manipulation, and Applications}
\author{Chih-Sung Chuu and Shengwang Du}
\institute{Chih-Sung Chuu \at Department of Physics and Frontier Research Center on
Fundamental and Applied Sciences of Matters, National Tsing Hua University, Hsinchu 30013, Taiwan, \email{cschuu@phys.nthu.edu.tw}
\and Shengwang Du \at Department of Physics, The Hong Kong University of Science and Technology, Clear Water Bay, Kowloon, Hong Kong, People's Republic of China, \email{dusw@ust.hk}}
%
%
\maketitle


\abstract{In this chapter, we review recent advances in generating narrowband biphotons with long coherence time using spontaneous parametric interaction in monolithic cavity with cluster effect as well as in cold atoms with electromagnetically induced transparency. Engineering and manipulating the temporal waveforms of these long biphotons provide efficient means for controlling light-matter quantum interaction at the single-photon level. We also review recent experiments using temporally long biphotons and single photons.
\newline\indent
}

\section{\label{sec:Introduction}Introduction}

Entangled photon pairs, termed \textit{\keyword{biphoton}s}, have been benchmark tools in the field of quantum optics for testing fundamental quantum mechanics as well as for developing applications in quantum information technology, including realization of the Einstein-Podolsky-Rosen paradox \cite{EPR, HowellPRL2004}, test of violation of Bell's inequality \cite{BellInequality, ClauserPRL1969, AspectPRL1981}, quantum cryptography and key distribution \cite{QKD-E91}, quantum teleportation \cite{BennettPRL1993, PanNature2012}, quantum computation \cite{QuantumComputation}, \textit{etc.} Traditional methods of producing \keyword{biphoton}s include spontaneous parametric down conversion (SPDC) \cite{HarrisPRL1967, BurnhamPRL1970, RubinPRA1994} and spontaneous four-wave mixing (\keyword{SFWM}) \cite{LiPRL2005, FanOL2005, CohenPRL2009} in nonlinear solid-state materials. Photons from these nonlinear processes in free space forward-wave configuration usually have broad bandwidth ($>$ THz) and short coherence time ($<$ ps) such that their temporal quantum waveform can not be directly resolved by existing commercial single-photon counters (which have a typical resolution of about ns).\\ 

These broadband \keyword{biphoton}s are not suitable for recently proposed protocols of long-distance quantum communication and quantum network based on photon-atom interaction \cite{DLCZ, QuantumNetwork}, because an efficient photon-atom quantum interface requires single flying photons having a bandwidth sufficiently narrower than the atomic resonance (typically in MHz). To reduce their bandwidth, optical cavity may be used for active filtering \cite{OuPRL1999, Kuklewicz06, PanPRL2008, Scholz09}. However, multiple cavity modes are resonated simultaneously due to the broad gain linewidth of the forward-wave parametric interaction and the \keyword{biphoton}s are generated in multiple longitudinal modes. Additional passive filters locked to the desired mode of the resonant cavity is thus necessary to obtain single mode output with reduced generation rate and increased complexity.\\

Phase-matching condition plays an important role in determining the photon bandwidth in these nonlinear process. The loose constraint of \keyword{phase matching} in the forward-wave configuration make it difficult to generate ultranarrow-bandwidth \keyword{biphoton}s. To overcome this problem, one may use the cluster effect in a monolithic cavity with double-pass pumping. An alternative solution is to take the backward-wave configuration where the produced paired photons propagate in opposing directions. In this chapter, we review recent advances in generating narrowband \keyword{biphoton}s with long coherence time using forward-wave and backward-wave parametric interaction. In particular, we describe the monolithic resonant down-conversion with cluster effect \cite{ChuuPRA2011, ChuuAPL2012} and the resonant \keyword{SFWM} in cold atoms with electromagnetically induced transparency (\keyword{EIT})\cite{Subnatural,DuOptica2014,DuJOSAB2008}.

\section{Monolithic Resonant Parametric Down-Conversion with Cluster Effect}
\label{sec:1}

A resonant parametric down-converter can generate single-mode biphotons without external filtering when its gain linewidth is narrower than the spacing of adjacent resonant modes. For a doubly resonator where the signal and idler fields are simultaneously resonant, the mode spacing is determined by the \keyword{cluster spacing} $\Delta \Omega_c$ between the simultaneous resonant signal the idler modes. The \keyword{cluster spacing} can be calculated by solving the following equations \cite{Eckardt91},
\begin{eqnarray}
\pm 1 &=& M(\omega) \Delta \Omega^2_c+N(\omega) \Delta \Omega_c, \nonumber \\
M(\omega) &=& [L/(2 \pi c)]\{2[n^{'}_s (\omega)+n^{'}_i(\omega_i)]+\omega_s n^{''}_s(\omega)+\omega_i n^{''}_i(\omega_i)\}, \nonumber \\
N(\omega) &=& [L/(\pi c)][n_s-n_i+\omega_s n^{'}_s (\omega_s)-\omega_i n^{'}_i (\omega_i)],
\end{eqnarray}
where $n_s$ and $n_i$ are the refractive indices at the signal and idler frequencies, and $n^{'}_{s,i}$ and $n^{''}_{s,i}$ are the first and second frequency derivatives, respectively.\\

If we assume that the idler mode spacing $\Delta_i$ is slightly less than the signal mode spacing $\Delta_s$ and the group velocity dispersion is negligible, the \keyword{cluster spacing} can also be obtained by multiplying the idler mode spacing by the number of idler modes between two doubly resonant modes,
\begin{equation}
\Delta \Omega_c \cong \frac{\Delta_s\ \Delta_i}{\Delta_s - \Delta_i}.
\label{eq:cluster}
\end{equation}
For small $\Delta_s - \Delta_i$, the \keyword{cluster spacing} will be much larger than the signal or idler mode spacing. Moreover, since $\Delta \Omega_c$ decreases with the length of cavity, a monolithic doubly resonator has the largest possible \keyword{cluster spacing} and can be used to reduce the number of resonant modes within the gain curve.\\

For parametric down-conversion of the forward wave type, a monolithic doubly resonator by itself is not enough for generating biphotons in a single longitudinal mode. This is because the gain linewidth of the parametric interaction is larger than the \keyword{cluster spacing}, which allows biphotons to be generated in multiple longitudinal modes. As an example, we consider a 3-cm long monolithic PPKTP crystal pumped by a continuous-wave 532 nm laser. At type-II phase matching, the gain linewidth $\Delta f_{sp} = 0.885 c/(|n_s^{(g)}-n_i^{(g)}|L) = 93.2$ GHz and the \keyword{cluster spacing} $\Delta \Omega_c = c/(2|n_s^{(g)}-n_i^{(g)}|L) = 52.6$ GHz where $n_{s,i}^{(g)}$ are the group indices at the signal and idler frequencies and $L$ is the crystal or cavity length. Because $\Delta f_{sp} > \Delta \Omega_c$, multimode biphotons are generated and external filters are required to eliminate biphotons in unwanted modes to obtain single-mode biphotons. This results in reduced biphoton generation rate in additional to increased complexity of the biphoton source.

\subsection{Single-mode output}

To obtain single-mode biphotons without additional filtering, it is necessary to increase the mode spacing and reduce the gain linewidth. This can be achieved by using \keyword{double-pass pumping}, for example, by depositing a high reflection coating at the pump frequency on the output face of the crystal. The \keyword{double-pass pumping} effectively doubles the length of the parametric interaction or reduces the gain linewidth by half. The broaden gain linewidth $\Delta f_{dp} = 0.443 c/(|n_s^{(g)}-n_i^{(g)}|L)$ is thus smaller than the \keyword{cluster spacing} $\Delta \Omega_c = c/(2|n_s^{(g)}-n_i^{(g)}|L)$. Consider the previous example with a double-pass pump. The gain linewidth $\Delta f_{dp} = 46.6$ GHz is now narrower than the \keyword{cluster spacing} $\Delta \Omega_c = 52.6$ GHz, so biphotons can be generated in a single longitudinal mode without the need of additional filtering.\\

\begin{figure}[htbp]
\centering\includegraphics[scale=.55]{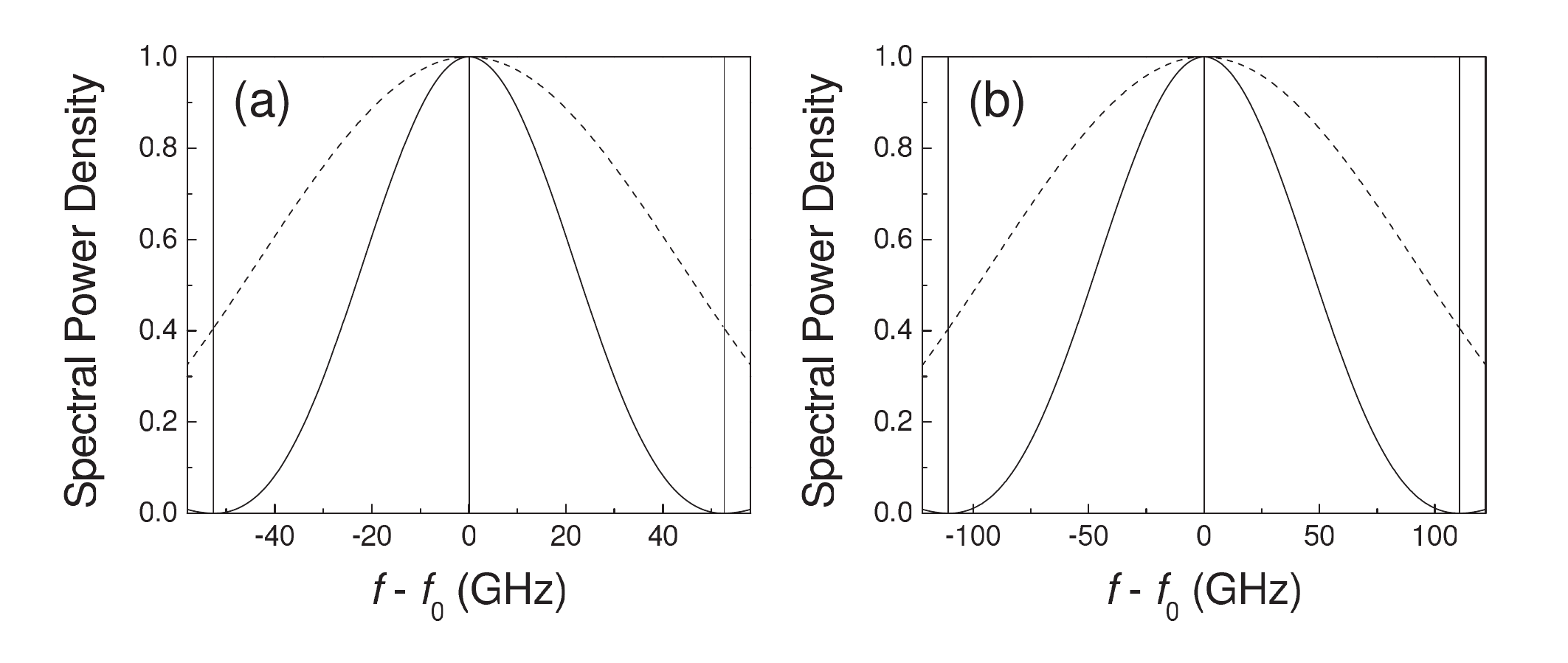}
\caption{\label{Chuu_fig1} Single-mode operation is shown as compared to multi-mode operation. The solid and dotted curves are the gain profiles (spectral power density) of a doubly resonant forward-wave parametric down-converters with double-pass and single-pass pumping, respectively. The vertical red lines represent the frequencies of the simultaneously resonant signal and idler pairs with one mode taken at the center frequency $f_0$ of the gain curve. Quasi-phasematching is chosen for (a) degenerate and (b) non-degenerate frequencies.}
\end{figure}

The principle of single-mode operation is illustrated in Fig.~\ref{Chuu_fig1}(a), where the solid curve is the gain profile and the vertical lines represent the frequencies of the simultaneously resonant signal and idler modes. With one resonant mode aligned to the center of the gain curve, the adjacent modes are outside the gain region. For comparison, the gain curve of a monolithic resonator with single-pass pumping is also shown as a dotted curve where more than one resonant signal and idler pairs are present within the gain curve.\\

As another example, we consider the generation of single-mode non-degenerate biphotons. We assume a 3-cm long monolithic PPKTP crystal pumped by a double-pass 525.5 nm laser. Such source may be useful for applications such as quantum repeater \cite{BriegelPRL1998}. For example, the signal photons at 795 nm can be stored in nearby atomic memories while idler photons at 1550 nm can be sent through a fiber to interfere on a distant beam splitter for creating entanglement between two remote locations. As shown in Fig.~\ref{Chuu_fig1}(b), the gain linewidth $\Delta f_{dp} = 98.1$ GHz is narrower than the \keyword{cluster spacing} $\Delta \Omega_c = 110.7$ GHz and the biphotons are generated in a single longitudinal mode.

\subsection{Experimental realization}

\begin{figure}
\centering
\includegraphics[scale=.9]{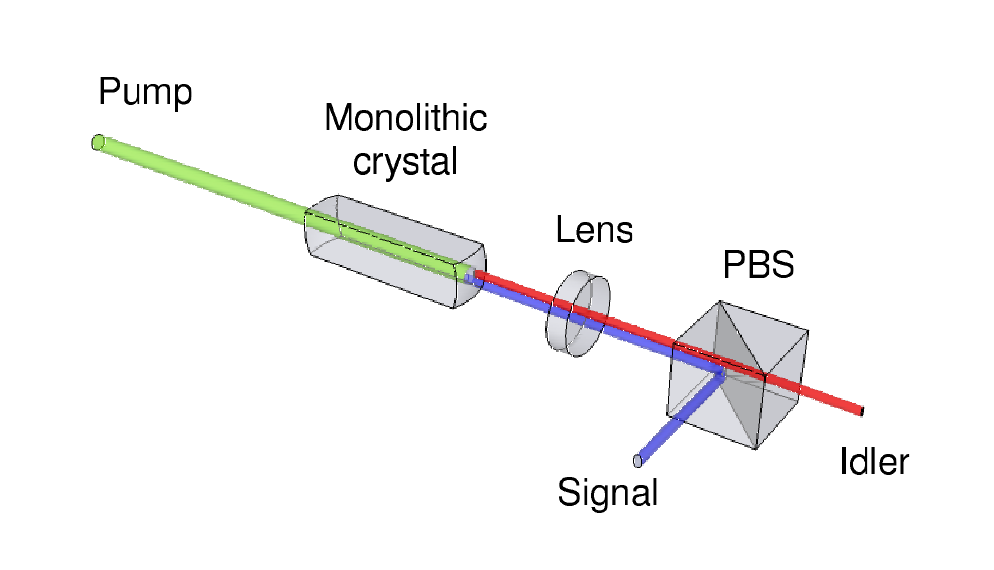}
\caption{\label{Chuu_fig2} A monolithic doubly resonant parametric down-converter for generating single-mode biphotons. The pump is double-passed by the end face of the monolithic crystal, which has a high reflection coating at the signal, idler, and pump wavelengths. The lens is used to collimate the signal and idler fields. The PBS (polarizing beam splitter) separates the signal and idler fields before they are detected by the single-photon detection modules (not shown).}
\end{figure}

Chuu, Yin, and Harris \cite{ChuuAPL2012} have demonstrated a miniature ultrabright source of narrowband biphotons using a monolithic resonant parametric down-converter without external filtering. A schematic of the experimental setup is shown in Fig.~\ref{Chuu_fig2}. The biphoton source uses a nonlinear crystal of which the end faces are spherically polished and deposited with a high reflection coating at the signal and idler wavelengths to form a monolithic cavity. The high reflection coating on one end face is also highly reflective at the pump wavelength for \keyword{double-pass pumping}. The combination of monolithic cavity, \keyword{double-pass pumping}, and type-II phase matching allows the generation of single-mode biphotons near degeneracy. When the pump is increased above threshold to produce parametric oscillation, the output is single-mode, as seen by a scanning Fabry Perot interferometer. This suggest that only a single cluster is resonant.\\

\begin{figure}
\centering
\includegraphics[scale=.9]{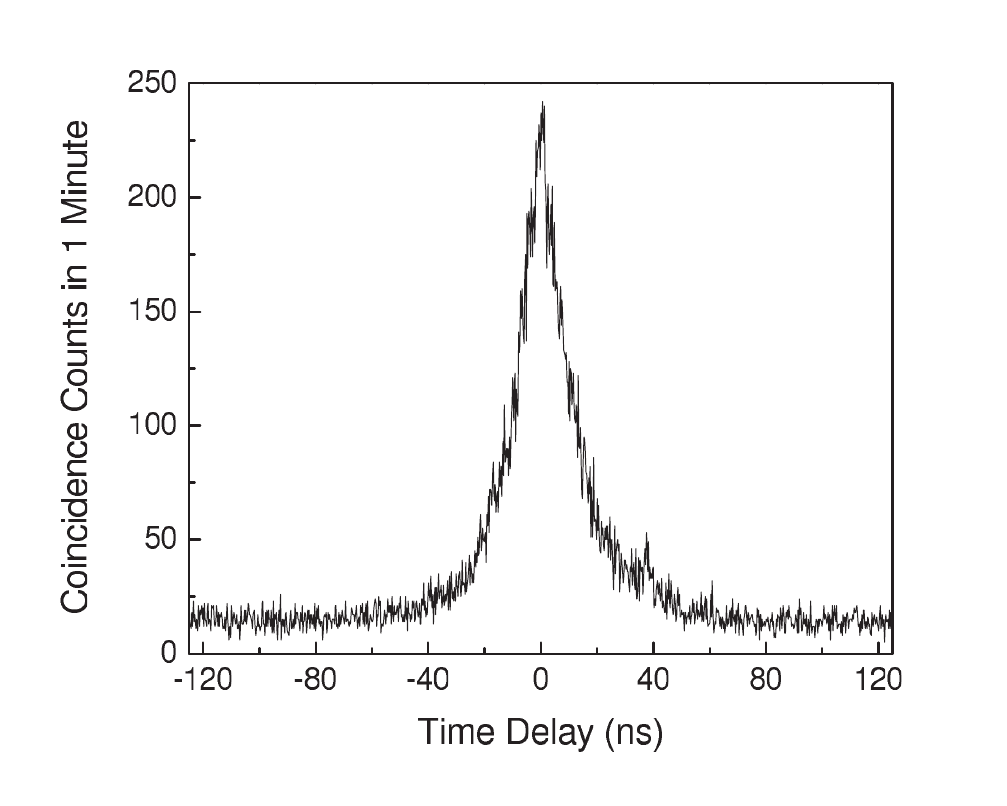}
\caption{\label{Chuu_fig3} Glauber correlation function of the biphotons. The coincidence counts are measured as a function of the time delay between the signal and idler photons at a pump power of 700 $\mu$W. The function has two exponential decays with decay constants of $13.29 \pm 0.14$~ns and $11.33 \pm 0.12$~ns for time delay greater than or less than zero. The biphoton correlation time is $17.07 \pm 0.13$~ns.}
\end{figure}

A typical coincidence measurement of the biphoton wavepacket is shown in Fig.~\ref{Chuu_fig3}. The measured curve shows two asymmetric exponential decays possibly due to the different reflectivity for orthogonal polarizations. The biphoton correlation time is found to be $T_c \cong 17$~ns and the biphoton bandwidth is $\Delta \omega \cong 2 \pi \cdot 8$~MHz. Correcting for the collection efficiency, the generation rate is $R=1.10 \times 10^5$ biphotons/(s\ mW) which gives a spectral brightness of $R/\Delta \omega = 1.34 \times 10^4$ biphotons/(s~MHz~mW).

\section{Backward-wave Biphoton Generation}

Parametric down-conversion of the \keyword{backward-wave} type provides an alternative way to realize an ultrabright biphoton source without external filtering. As compared to conventional forward-wave schemes, such as spontaneous parametric down conversion (SPDC) in nonlinear crystals, the backward geometry allows a much tighter constraint of \keyword{phase matching} that leads to substantially narrow linewidth.

\subsection{\label{sec:General Formulism}General Formulism: Free space}

\begin{figure}[h]
\includegraphics[width=0.85\linewidth]{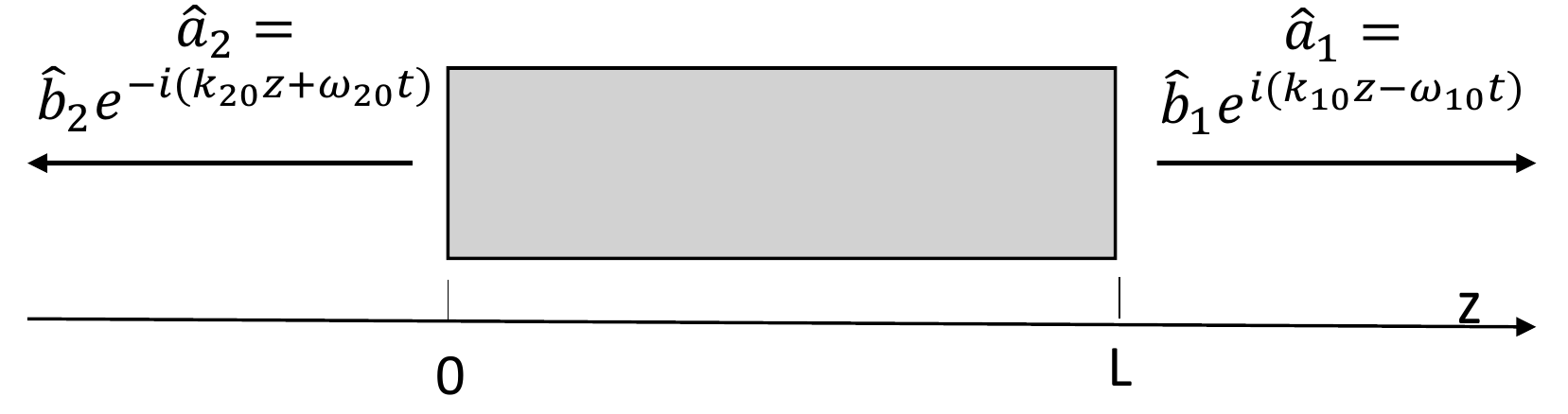}
\centering
\caption{\label{Chuu_fig4}Schematics of \keyword{backward-wave} \keyword{biphoton} generation from a nonlinearly driven medium in free space.}
\end{figure}

The \keyword{backward-wave} \keyword{biphoton} generation in free space is schematically illustrated in Fig.~\ref{Chuu_fig4}, where the generated field $\hat{a}_1$ propagates along the +z direction and $\hat{a}_2$ along the -z direction. For SPDC with $\chi^{(2)}$ nonlinearity, this geometry can be achieved by quasiphase matching \cite{Byer1992} to submicron periodicity and driven by a single pump laser beam \cite{CanaliasAPL2003, CanaliasAPL2005, CanaliasNP2007}. For \keyword{SFWM} driven by $\chi^{(3)}$ nonlinearity, the \keyword{phase matching} condition can be satisfied by aligning two coherent driving laser beams \cite{BalicPRL2005, DuJOSAB2008}. As compared to conventional forward-wave schemes, the \keyword{backward-wave} geometry allows a much tighter constraint of \keyword{phase matching} that leads to substantially narrow linewidth.

We describe the spontaneously generated paired photons as quantized field operators
\begin{eqnarray}
\hat{a}_1(z,t)&=&\hat{b}_1(z,t)e^{i(k_{10}z-\omega_{10}t)},\nonumber \\
\hat{a}_2(z,t)&=&\hat{b}_2(z,t)e^{-i(k_{20}z+\omega_{20}t)},
\label{eq:FieldOperatorAZT}
\end{eqnarray}
where $\omega_{i0}$ are the central angular frequencies and $k_{i0}=\omega_{i0}/c$ are the wave numbers in vacuum. $\hat{b}_i(z,t)$ are their slowly varying envelopes and follow the Fourier transform
\begin{eqnarray}
\hat{b}_i(z,t)=\frac{1}{\sqrt{2\pi}}\int\hat{b}_i(z,\omega)e^{-i\omega t}d\omega.
\label{eq:FieldOperatorbZT}
\end{eqnarray}
The frequency domain field operators are governed by the following Heisenberg-Langevin coupled equations \cite{KolchinPRA2007}:
\begin{eqnarray}
\left[\frac{\partial}{\partial z}+ \alpha_{1}(\omega)-i\frac{\Delta k_{0}}{2}\right]\hat{b}_{1}(z,\omega)=\kappa_1(\omega) \hat{b}^{\dag}_{2}(z,-\omega)+\hat{F}_1(z,\omega), \nonumber \\
\left[\frac{\partial}{\partial z}+ g_{2}(\omega)+i\frac{\Delta k_{0}}{2}\right]\hat{b}^{\dag}_{2}(z,-\omega)=\kappa_2(\omega) \hat{b}_{1}(z,\omega)+\hat{F}^{\dag}_2(z,-\omega),
\label{eq:CoupledEqF}
\end{eqnarray}
where $\alpha_1(\omega)=-i\frac{\omega_{10}}{2c}\chi_{1}(\omega_{10}+\omega)$ and $g_2(\omega)=-i\frac{\omega_{10}}{2c}\chi_{2}^{*}(\omega_{20}-\omega)$ describe the linear propagation effects with the linear susceptibilities $\chi_i$, $\kappa_i(\omega)$ are the nonlinear parametric coupling coefficients, $\Delta k_0$ is the phase mismatching in vacuum, and $\hat{F}_i(z,\omega)$ are the Langevin noise operators. Neglecting the Langenvin noise operators that contribute to uncorrelated noise photons, the above equations reduce to
\begin{eqnarray}
\left[\frac{\partial}{\partial z}+ \alpha_{1}(\omega)-i\frac{\Delta k_{0}}{2}\right]\hat{b}_{1}(z,\omega)=\kappa_1(\omega) \hat{b}^{\dag}_{2}(z,-\omega), \nonumber \\
\left[\frac{\partial}{\partial z}+ g_{2}(\omega)+i\frac{\Delta k_{0}}{2}\right]\hat{b}^{\dag}_{2}(z,-\omega)=\kappa_2(\omega) \hat{b}_{1}(z,\omega),
\label{eq:CoupledEq}
\end{eqnarray}
which are subject to the boundary conditions $[\hat{b}_1(0,\omega),\hat{b}^{\dag}_1(0,\omega')]=[\hat{b}_2(L,\omega),\hat{b}^{\dag}_2(L,\omega')]=\delta(\omega-\omega')$ and $\langle\hat{b}^{\dag}_1(0,\omega')\hat{b}_1(0,\omega)\rangle=\langle\hat{b}^{\dag}_2(L,\omega')\hat{b}_2(L,\omega)\rangle=0$. The general solution to Eq.~(\ref{eq:CoupledEq}) is given as
\begin{eqnarray}
\hat{b}_{1}(L,\omega)=A(\omega)\hat{b}_{1}(0,\omega)+B(\omega)\hat{b}^{\dag}_{2}(L,-\omega), \nonumber \\
\hat{b}^{\dag}_{2}(L,-\omega)=C(\omega)\hat{b}_{1}(0,\omega)+D(\omega)\hat{b}^{\dag}_{2}(L,-\omega),
\label{eq:Solution}
\end{eqnarray}
where
\begin{eqnarray}
A(\omega)&=&\frac{Q e^{-(\alpha_1+g_2)L/2}}{q \sinh(QL/2)+Q\cosh(QL/2)}, \nonumber \\
B(\omega)&=&\frac{2 \kappa_1}{q+Q\coth(QL/2)}, \nonumber \\
C(\omega)&=&\frac{-2 \kappa_2}{q+Q\coth(QL/2)},\nonumber \\
D(\omega)&=&\frac{Q e^{(\alpha_1+g_2)L/2}}{q \sinh(QL/2)+Q\cosh(QL/2)}.
\label{eq:ABCD}
\end{eqnarray}
Here $q(\omega)=\alpha_1(\omega)-g_2(\omega)-i\Delta k_0$ and $Q(\omega)=\sqrt{q^2(\omega)+4\kappa_1(\omega)\kappa_2(\omega)}$. The single photon rates can be calculated from
\begin{eqnarray}
R_1=\langle \hat{b}_1^{\dag}(L,t)\hat{b}_1(L,t)\rangle=\frac{1}{2\pi}\int|B(\omega)|^2 d\omega, \nonumber\\
R_2=\langle \hat{b}_2^{\dag}(0,t)\hat{b}_2(0,t)\rangle=\frac{1}{2\pi}\int|C(\omega)|^2 d\omega.
\label{eq:PhotonRates}
\end{eqnarray}
The two-photon Glauber correlation function can be determined by
\begin{eqnarray}
G_{12}^{(2)}(\tau)=\langle \hat{b}_1^{\dag}(L,t+\tau)\hat{b}_2^{\dag}(0,t)\hat{b}_2(0,t)\hat{b}_1(L,t+\tau) \rangle =|\psi(\tau)|^2+R_1 R_2,
\label{eq:GlauberCorrelation}
\end{eqnarray}
where the two-photon wave function is
\begin{eqnarray}
\psi(\tau)=\langle \hat{b}_2(0,t)\hat{b}_1(L,t+\tau) \rangle =\frac{1}{2\pi}\int B(\omega)D^*(\omega)e^{-i\omega\tau}d\omega.
\label{eq:BiphotonWaveFunction}
\end{eqnarray}
Alternatively, the two-photon wave function can also be obtained from
\begin{eqnarray}
\psi(\tau)=\langle \hat{b}_1(L,t+\tau)\hat{b}_2(0,t) \rangle =\frac{1}{2\pi}\int A(\omega)C^*(\omega)e^{-i\omega\tau}d\omega.
\label{eq:BiphotonWaveFunction2}
\end{eqnarray}
When the system is conservative and the commutation relation of the field operators is preserved, Eqs.~(\ref{eq:BiphotonWaveFunction}) and (\ref{eq:BiphotonWaveFunction2}) are equivalent. When the Langevin noise fluctuations exist, Eqs.~(\ref{eq:BiphotonWaveFunction}) and (\ref{eq:BiphotonWaveFunction2}) become approximated solutions \cite{DuShapingBiphoton2014}. The photon pair generation rate can be calculated from
\begin{eqnarray}
R=\int|\psi(\tau)|^2d\tau.
\label{eq:PhotonPairrate}
\end{eqnarray}
The normalized two-photon cross-correlation function is
\begin{eqnarray}
g_{12}^{(2)}(\tau)=\frac{G_{12}^{(2)}(\tau)}{R_1 R_2}=1+\frac{|\psi(\tau)|^2}{R_1 R_2}.
\label{eq:g12}
\end{eqnarray}

It is instructive to see what parameters play important role in determining the two-photon wave function under some reasonable approximations. To ensure the spontaneous photon generation remains below the threshold and muti-photon-pair events are suppressed, we work in the low parametric gain regime, \textit{i.e.}, $|\kappa_1(\omega)\kappa_2(\omega)|\ll |q^2|$. We further take $\kappa_1\simeq\kappa_2\simeq\kappa$ and assume the linear loss and gain of the medium to the generated fields can be neglected for the generated fields (\textit{i.e.}, $\chi_i=\chi_i^*$). Under these conditions, the ABCD parameters in Eq.(\ref{eq:ABCD}) can be written approximately
\begin{eqnarray}
A&\simeq & e^{i\Delta k_{1}L}e^{i\Delta k_0 L/2}, \nonumber \\
B&\simeq & \kappa L \textrm{sinc}(\Delta k L/2) e^{i\Delta k L/2}, \nonumber \\
C&\simeq & -\kappa L \textrm{sinc}(\Delta k L/2) e^{i\Delta k L/2},\nonumber \\
D&\simeq & e^{-i\Delta k_{2}L}e^{i\Delta k_0 L/2},
\label{eq:ABCDapp}
\end{eqnarray}
where $\Delta k_{i}=\chi_i\omega_{i0}/(2c)$ and $\Delta k =\Delta k_{1}-\Delta k_{2}+\Delta k_0$. The \keyword{biphoton} wave function in Eq.~(\ref{eq:BiphotonWaveFunction}) then becomes
\begin{eqnarray}
\psi(\tau)=\frac{L}{2\pi}\int \kappa(\omega)\Phi(\omega)e^{-i\omega\tau}d\omega,
\label{eq:BiphotonWaveFunction1}
\end{eqnarray}
where $ \Phi(\omega)$ is the the longitudinal detuning function
\begin{eqnarray}
\Phi(\omega)=\textrm{sinc} \left [\frac{\Delta k(\omega)L}{2}\right ]e^{i[\Delta k_{1}(\omega)+\Delta k_{2}(\omega)]L/2}.
\label{eq:longitudinaldetuningfunction}
\end{eqnarray}
Equations (\ref{eq:BiphotonWaveFunction1}) and (\ref{eq:longitudinaldetuningfunction}) are the same as those obtained following a perturbation treatment in the interaction picture \cite{DuJOSAB2008}. The \keyword{biphoton} joint spectrum is determined by two factors: the nonlinear parametric coupling coefficient $\kappa(\omega)$ and the longitudinal detuning function $\Phi(\omega)$ from the linear phase-matching and propagation effects.

\subsection{General Formalism: Resonant SPDC}

Doubly resonant optical cavity plays an important role in obtaining single-mode output for ultrabright biphoton generation. Here we develop the theory of resonant biphoton generation in the Heisenberg picture.\\

We denote the signal and idler fields internal to the cavity by the standing-wave cavity operators
\begin{eqnarray}
a_s(t,z) &=& b_s(t) \exp(-i \Omega_q t) \sin(k_q z) \nonumber \\
a_i(t,z) &=& b_i(t) \exp(-i \Omega_r t) \sin(k_r z)
\end{eqnarray}
where $b_s(t)$ and $b_i(t)$ are the slowly varying envelopes, $\Omega_q$ and $\Omega_r$ are the cold cavity frequencies, $k_q = q \pi /L$ and $k_r = r \pi /L$. We assume that only one pair of signal and idler fields is resonant with the $q$th and $r$th cavity modes simultaneously (we will justify this in the following section). Since only the components nonorthogonal to the $q$th and $r$th cavity mode interact with the signal and idler fields respectively, the generated dipole moment operators for the signal and idler fields, which are proportional to $b_i(t) \exp(-i \Omega_q t) \exp(i k_q z) \exp(i \Delta k z)$ and $b_s(t) \exp(-i \Omega_r t) \exp(i k_r z) \exp(i \Delta k z)$ where the phase mismatch $\Delta k = k_p - k_r - k_q$, are projected against $\sin(k_q z)$ and $\sin(k_r z)$.\\

Using the input-output coupling formalism \cite{CollettPRA1984, SensarnPRL2009}, the equations for the evolution of $b_s(t)$ and $b_i(t)$ and their relation to the incident fields are
\begin{eqnarray}
\frac{\partial b_s (t)}{\partial t} + \frac{\Gamma_s}{2} b_s (t) &=& - i \kappa\ b^{\dagger}_i(t) + \sqrt{\gamma_s}\ b^{\rm in}_s(t) \nonumber \\
\frac{\partial b^{\dagger}_i (t)}{\partial t} + \frac{\Gamma_i}{2} b^{\dagger}_i (t) &=& i \kappa\ b_s(t) + \sqrt{\gamma_i}\ b^{\rm in \dagger}_i(t), \label{eq:coupled1}
\end{eqnarray}
where $ b^{\rm in}_s (t)$ and $ b^{\rm in}_i (t)$ are the fields incident on the resonant cavity and $\kappa$ is the coupling constant. With $\{\Delta_s, \Delta_i \}$ and $\{ r_s, r_i \}$ denoting the spacing of the cavity modes and the mirror reflectivity, respectively, the output coupling rates of the signal and idler fields are $\gamma_s = \Delta_s (1-r_s)$ and $\gamma_i = \Delta_i (1-r_i)$. With $\xi_s$ and $\xi_i$ defined as the single-pass power loss for the signal and idler fields in the crystal, the total cavity decay rates are $\Gamma_s=2 \xi_s \Delta_s + \gamma_s$ and $\Gamma_i=2 \xi_i \Delta_i + \gamma_i$.\\

The slowly varying output fields $b^{\rm out}_s(t)$ and $b^{\rm out \dagger}_i(t) $ are related to the internal and incident fields by
\begin{eqnarray}
b^{\rm out}_s(t) &=& \sqrt{\gamma_s}\ b_s(t) - b^{\rm in}_s(t) \nonumber \\
b^{\rm out \dagger}_i(t) &=& \sqrt{\gamma_i}\ b^{\dagger}_i(t) - b^{\rm in \dagger}_i(t).
\label{eq:coupled2}
\end{eqnarray}
They can be solved by transforming the coupled equations to the frequency domain with the Fourier pair
\begin{eqnarray}
b(t) &=& \int_{- \infty}^{\infty} b(\omega') \exp (-i \omega' t) d \omega' \nonumber \\
b(\omega) &=& \frac{1}{2 \pi} \int_{- \infty}^{\infty} b(t') \exp (i \omega t') d t'
\end{eqnarray}
and converting to rapidly varying quantities
\begin{eqnarray}
a(\omega_{s,i}) &=& b(\omega_{s,i}-\Omega_{q,r}) \nonumber \\
a^{\dagger}(\omega_{s,i}) &=& b^{\dagger}(\omega_{s,i}+\Omega_{q,r})
\end{eqnarray}
The output fields $a^{\rm out}_s (\omega)$ and $a^{\rm out \dagger}_i (-\omega_i)$ can then be expressed in terms of incident fields $a^{\rm in}_s (\omega)$ and $a^{\rm in \dagger}_i (-\omega_i)$,
\begin{eqnarray}
a^{\rm out}_s (\omega) &=& A(\omega)\ a^{\rm in}_s (\omega) + B(\omega)\ a^{\rm in \dagger}_i (- \omega_i) \nonumber \\
a^{\rm out \dagger}_i (- \omega_i) &=& C(\omega)\ a^{\rm in}_s (\omega) + D(\omega)\ a^{\rm in \dagger}_i (- \omega_i), \label{eq:ABCD1}
\end{eqnarray}
with commutators $[a^{\rm in}_j(\omega_1),a^{\rm in \dagger}_k(\omega_2)]=[a^{\rm out}_j(\omega_1),a^{\rm out \dagger}_k(\omega_2)]=\frac{1}{2\pi}\delta_{jk}\delta(\omega_1-\omega_2)$. For small gain the coefficients are given by
\begin{eqnarray}
A(\omega) &=& \frac{\gamma_s - \Gamma_s/2+i (\omega-\Omega_q)}{\Gamma_s/2-i (\omega-\Omega_q)} \nonumber\\
B(\omega) &=& \frac{-i \kappa \sqrt{\gamma_s \gamma_i}}{[\Gamma_s/2-i (\omega-\Omega_q)][\Gamma_i/2+i (\omega_i-\Omega_r)]} \nonumber \\
C(\omega) &=& \frac{i \kappa \sqrt{\gamma_s \gamma_i}}{[\Gamma_s/2-i (\omega-\Omega_q)][\Gamma_i/2+i (\omega_i-\Omega_r)]} \nonumber \\
D(\omega) &=& \frac{\gamma_i - \Gamma_i/2-i (\omega_i-\Omega_r)}{\Gamma_i/2+i (\omega_i-\Omega_r)} \label{eq:ABCD2}
\end{eqnarray}\\

With Eq.~(\ref{eq:ABCD1}) and (\ref{eq:ABCD2}), we obtain the spectral power density at the signal frequency
\begin{equation}
S(\omega) = \frac{1}{2 \pi}\left| B(\omega) \right|^2 = \frac{8 \gamma_s \gamma_i \kappa^2}{\pi [4(\omega-\Omega_q)^2 + \Gamma_s^2][4(\omega_i-\Omega_r)^2 + \Gamma_i^2]},
\label{eq:spectrum}
\end{equation}
with the \keyword{biphoton} linewidth $\Delta \omega = [(\sqrt{\Gamma_s^4+6 \Gamma^2_s \Gamma^2_i + \Gamma_i^4}-\Gamma^2_s-\Gamma^2_i)/2]^{1/2}$, and the total paired count rate at exact \keyword{phase matching},
\begin{equation}
R = \frac{1}{2 \pi} \int_{-\infty}^{\infty} |B(\omega')|^2 d \omega' = \frac{4 \gamma_s \gamma_i \kappa^2}{\Gamma_s \Gamma_i (\Gamma_s+\Gamma_i)}.
\end{equation}
Compared to a non-resonant forward-wave SPDC of the same crystal length and pumping power [Eq.~(\ref{eq:PhotonPairrate})], the generation rate of a lossless resonant backward-wave SPDC is thus increased by a factor of
\begin{equation}
\eta_r \approx \frac{8 \cal{F}}{\pi r^{1/2}} \frac{|v_s-v_i|}{(v_s+v_i)},
\end{equation}
where ${\cal{F}}$ is the cavity finesse, $r$ is the mirror reflectivity, and $v_{s,i}$ are the group velocities of the signal and idler photons. The spectral brightness is increased by a factor of $\eta_b = \eta_r \Delta \omega_G/\Delta \omega$ with $\Delta \omega_G$ being the gain linewidth of the parametric interaction.\\

If the generation rate of \keyword{biphoton}s is small as compared to the inverse of the temporal length of the \keyword{biphoton}, the accidental two-photon events may be neglected and the time domain Glauber correlation function (namely the \keyword{biphoton} wavepacket) is given by \cite{Harris07}
\begin{eqnarray}
G^{(2)}(\tau) &=& \left| \frac{1}{2 \pi}\int_{-\infty}^{\infty} A(\omega')C^*(\omega') e^{i \omega' \tau} {\rm d}\omega' \right|^2 + \left| \frac{1}{2 \pi} \int_{-\infty}^{\infty} |B(\omega')|^2 d \omega' \right|^2 \nonumber \\
 &\approx& \frac{4 \Gamma_s \Gamma_i \kappa^2}{(\Gamma_s + \Gamma_i)^2} \times
\left\{ \begin{array}{ll}
 e^{\Gamma_s \tau} &\mbox{, $\tau<0$ } \\
 e^{-\Gamma_i \tau} &\mbox{, $\tau>0$ }.
       \end{array} \right.
\label{eq:fullG2}
\end{eqnarray}
where $\tau=t_i-t_s$ is the time delay between the arrival of the signal and idler photons. The \keyword{biphoton} correlation time is then $T_c = (\ln 2) (1/\Gamma_s+1/\Gamma_i)$. The asymmetry of \keyword{biphoton} wavepacket in $\tau$ is due to the order of detection of the signal and idler photons.

\subsection{Single-mode output}

\begin{figure}
\centering
\includegraphics[scale=0.45]{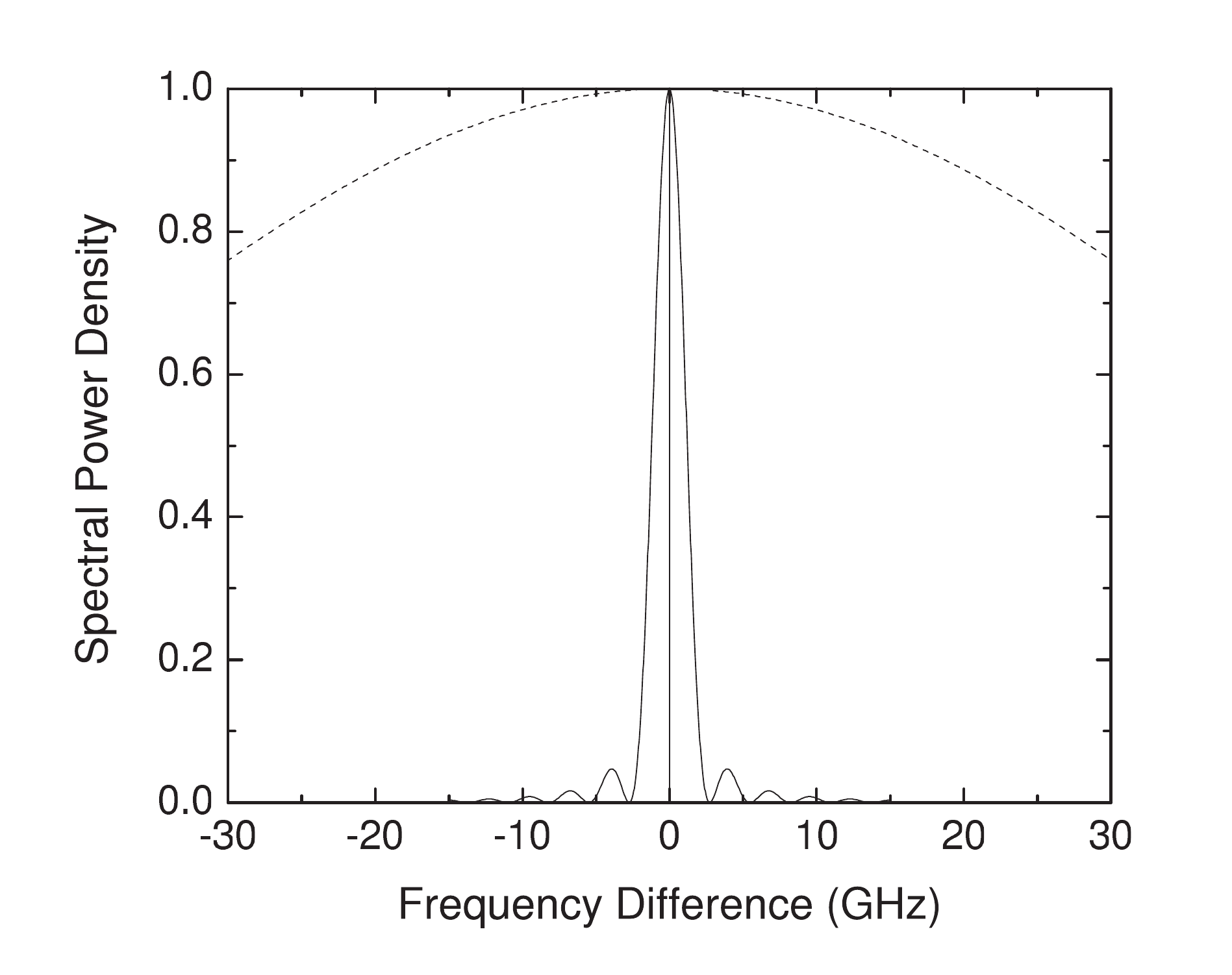}
\caption{\label{Chuu_fig5} Gain profile of \keyword{backward-wave} parametric interaction (solid) as compared to that of a forward-wave interaction (dash). The vertical line at 0 denotes a mode pair of the resonant cavity that is taken at the degenerate frequency. The next mode pair, separated by the \keyword{cluster spacing} is outside the gain linewidth of the \keyword{backward-wave} parametric interaction.}
\end{figure}

The narrow gain linewidth of the \keyword{backward-wave} parametric interaction is advantageous for generating single-mode \keyword{biphoton}s in a monolithic doubly resonant down-converter. Consider a 3-cm long PPKTP crystal placed inside a cavity of the same length. The \keyword{cluster spacing} $\Delta \Omega_{\rm Cl} \cong 2 \pi \cdot 1.75$ cm$^{-1}$ and the gain linewidth of the parametric interaction $\Delta \omega_G \cong 2 \pi \cdot 0.08$ cm$^{-1}$. Since $\Delta \Omega_{\rm Cl}$ is much broader than $\Delta \omega_G$, there will only be one doubly resonant mode within the gain linewidth when the cavity is properly tuned. This is shown in Fig.~\ref{Chuu_fig5} where the gain curves of the forward-wave type (red) is also shown for comparison.\\

\begin{figure}
\centering
\includegraphics[scale=.75]{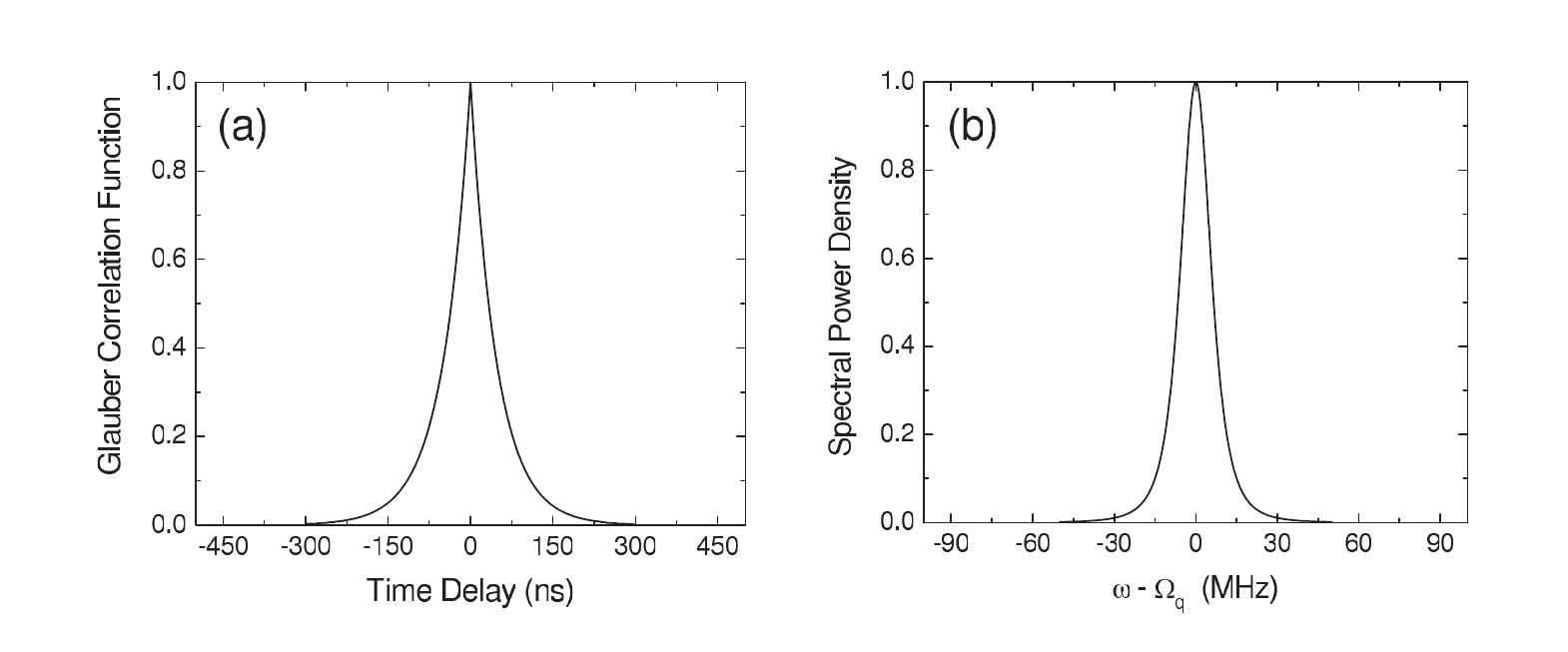}
\caption{\label{Chuu_fig6}(a) Time domain \keyword{biphoton} wavepacket of the \keyword{backward-wave} spontaneous down-conversion within a resonant cavity. (b) Spectral power density at the signal frequency.}
\end{figure}

The properties of the single-mode \keyword{biphoton}s can be calculated as for the case of forward-wave interaction. We assume a cavity finesse of 1000. The time-domain \keyword{biphoton} wavepacket is plotted in Fig.~\ref{Chuu_fig6}(a), which is slightly asymmetric due to the different ring-down times at the signal and idler frequencies. The \keyword{biphoton} correlation time is 68 ns. The spectral power density of the \keyword{biphoton}s is plotted in Fig.~\ref{Chuu_fig6}(b) and has a Lorentzian shape. The \keyword{biphoton} linewidth is $2 \pi \cdot 2.1$~MHz. For exact \keyword{phase matching}, the spectral brightness of the generated \keyword{biphoton}s is $8.16 \cdot 10^4$~s$^{-1}$MHz$^{-1}$ per mW of pump power and, as compared to a non-resonant forward-wave source of the same material, pumping power, and length, is about 80 000 times higher.

\subsection{Experimental challenge}

The \keyword{backward-wave} parametric interaction was proposed in the 1960's for implementing a mirrorless oscillator in the infrared regime \cite{Harris66}. The experimental demonstration, however, was not realized until forty years later \cite{CanaliasNP2007} because of the required quasi-\keyword{phase matching} with sub-micron periodicity \cite{CanaliasAPL2003,CanaliasAPL2005}.\\

To construct a \keyword{backward-wave} \keyword{biphoton} source as described above, a 532 nm laser may be used as the pump source to generate \keyword{biphoton}s at the degenerate frequency of 1.064 $\mu$m. For type-II \keyword{phase matching}, a KTP crystal periodically poled with a periodicity of $\Lambda=872$ nm is required to accomplish the third-order ($m=3$) quasi phasematching so that $k_p = K_G + k_s - k_i$, where the lattice k-vector $K_G=2 \pi m / \Lambda$. Realization of such source thus needs a KTP crystal that is periodically poled with a sub-micron periodicity. Although challenging, it can be done with current structuring technique \cite{CanaliasNP2007}.

\section{\label{sec:SFWM-EIT} Spontaneous Four-Wave Mixing with Electromagnetically Induced Transparency}

\begin{figure}
\includegraphics[width=0.85\linewidth]{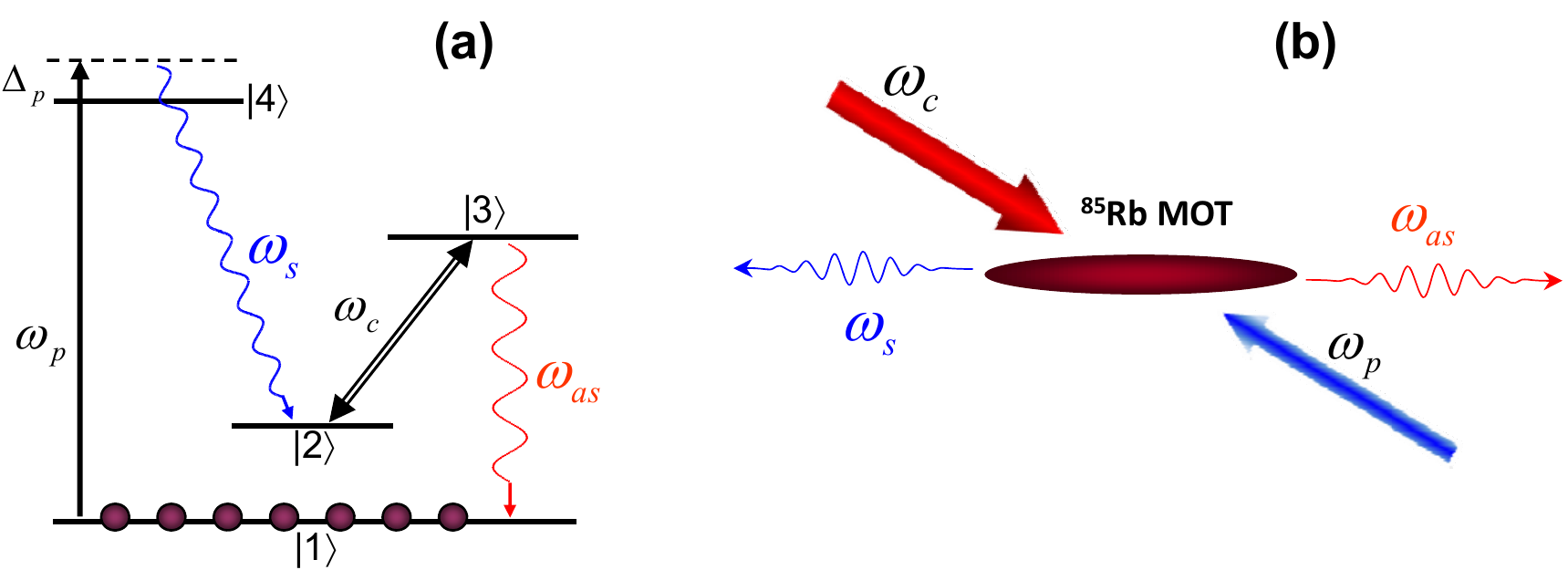}
\centering
\caption{\label{Chuu_fig7}Biphoton generation via \keyword{SFWM} in a four-level double-$\Lambda$ atomic system. (a) The atomic energy level diagram. (b) The backward-wave \keyword{biphoton} generation geometry.}
\end{figure}

Here we describe narrowband \keyword{biphoton} generation via \keyword{SFWM} with \keyword{EIT} \cite{EITHarris, MFleischhauer} in cold atoms. The modeled four-level double-$\Lambda$ atomic system and backward-wave configuration are shown in Fig.~\ref{Chuu_fig7}. $|1\rangle$ and $|2\rangle$ are two long-lived ground states (such as the two hyperfine ground states of alkali atoms) and between them there is no electric dipole transition. $|3\rangle$ and $|4\rangle$ are two excited states (in some scheme they can be the same state). A pump laser ($\omega_p$) excites atoms at the transition $|1\rangle\rightarrow|4\rangle$ with a detuning $\Delta_p$. A coupling laser ($\omega_c$) dresses the states $|2\rangle$ and $|3\rangle$ resonantly. In the presence of continuous-wave counter-propagating pump and coupling lasers, phase-matched backward paired Stokes ($\omega_s$) and anti-Stokes ($\omega_{as}$) photons are spontaneously produced following the transitions $|4\rangle\rightarrow|2\rangle$ and $|3\rangle\rightarrow|1\rangle$, respectively. We assume the pump laser is far detuned from the transition $|1\rangle\rightarrow|4\rangle$ and its excitation is weak such that the majority of atomic population remains in the ground state $|1\rangle$. Under this ground-state approximation, we further assume that both the pump and coupling laser beams are undepleted in the atomic medium. As shown in \ref{Chuu_fig7}(a), the coupling laser and the weak generated anti-Stokes field form a standard three-level $\Lambda$ \keyword{EIT} scheme. As a result, the coupling laser not only participates in the \keyword{SFWM} nonlinear process but also renders the medium transparent for the resonantly generated anti-Stokes photons. This \keyword{EIT} resonance dramatically enhances the \keyword{SFWM} nonlinear photon conversion efficiency. Moreover, in this system, the anti-Stokes photons propagate with a slow group velocity due to the \keyword{EIT} effect \cite{slowlight} while the Stokes photons travel nearly with the speed of light in vacuum. Below we show that this group velocity mismatching is the key to manipulating the phase-matching spectrum and thus the \keyword{biphoton} bandwidth. In the backward-wave configuration paired Stokes and anti-Stokes photons can propagate collinearly with the pump and coupling beams, and can also propagate in a right-angle geometry, depending on how well the relationship $\vec{k}_p+\vec{k}_c=0$ is satisfied, where $\vec{k}_{p,c}$ are wave vectors of the pump and coupling fields.

The cold atoms (without Doppler broadening effect) are confined within a long, thin cylindrical volume of a length $L$ and atomic density is $N$. The experiments reviewed in this chapter are mostly done with cold atoms in a two-dimensional (2D) magneto-optical trap (MOT) \cite{2DMOT}. To adapt the theory in Sec.~\ref{sec:General Formulism}, we replace the field index $1$ with $as$ and $2$ with $s$. The nonlinear parametric coupling coefficients ($\kappa_{as}$ and $\kappa_{s}$) for the \keyword{SFWM} process are connected to the third-order nonlinear susceptibilities ($\chi_{as}^{(3)}$ and $\chi_{s}^{(3)}$):
\begin{eqnarray}
\kappa_{as}(\omega)=i\frac{\omega_{as0}}{2c}\chi_{as}^{(3)}(\omega_{as0}+\omega)E_pE_c,\nonumber \\
\kappa_{s}(\omega)=i\frac{\omega_{s0}}{2c}\chi_{s}^{(3)*}(\omega_{as0}-\omega)E_p^*E_c^*.
\label{eq:kappa}
\end{eqnarray}
Here $E_p$ and $E_c$ are the electric field amplitude of the pump and coupling laser beams, resectively. Under the ground-state approximation, the third-order nonlinear susceptibility for the generated anti-Stokes and Stokes fields are
\begin{eqnarray}
\chi_{as}^{(3)}(\omega_{as0}+\omega)=\frac{N\mu_{13}\mu_{32}\mu_{24}\mu_{41}/(\varepsilon_0\hbar^3)}{(\Delta_{p}+i\gamma_{14})[|
\Omega_c|^2-4(\omega+i\gamma_{13})(\omega+i\gamma_{12})]}\nonumber \\
=\frac{-N\mu_{13}\mu_{32}\mu_{24}\mu_{41}/(\varepsilon_0\hbar^3)}{4(\Delta_{p}+i\gamma_{14})(\omega-\Omega_e/2+i
\gamma_{e})(\omega+\Omega_e/2+i\gamma_{e})},\nonumber \\\label{eq:Chi3as}
\end{eqnarray}
and
\begin{eqnarray}
\chi_{s}^{(3)}(\omega_{as0}-\omega)=\frac{N\mu_{13}\mu_{32}\mu_{24}\mu_{41}/(\varepsilon_0\hbar^3)}{(\Delta_{p}+i\gamma_{14})[|
\Omega_c|^2-4(\omega-i\gamma_{13})(\omega-i\gamma_{12})]}\nonumber \\
=\frac{-N\mu_{13}\mu_{32}\mu_{24}\mu_{41}/(\varepsilon_0\hbar^3)}{4(\Delta_{p}+i\gamma_{14})(\omega-\Omega_e/2-i
\gamma_{e})(\omega+\Omega_e/2-i\gamma_{e})}.\nonumber \\\label{eq:Chi3s}
\end{eqnarray}
$\mu_{ij}$ are the electric dipole matrix elements, $\Omega_c=\mu_{23}E_c/\hbar$ is the coupling Rabi frequency, and $\gamma_{ij}$ are dephasing rates, respectively. $\Delta_p=\omega_p-\omega_{41}$ is the pump detuning from the atomic transition $|1\rangle\rightarrow|4\rangle$. $\omega=\omega_{as}-\omega_{as0}$ is the detuning of the anti-Stokes photons from the transition $|1\rangle\rightarrow|3\rangle$, and we take $\omega_{as0}=\omega_{31}$ as the anti-Stokes central frequency. From Eqs.~(\ref{eq:Chi3as}) and (\ref{eq:Chi3s}), we have $\chi_{as}^{(3)}(\omega_{as0}+\omega)=\chi_{s}^{(3)}(\omega_{as0}+\omega)$. As the pump laser is far detuned ($\Delta_p\gg\gamma_{14}$), we further obtain $\chi_{as}^{(3)}(\omega_{as0}+\omega)\simeq\chi_{s}^{(3)*}(\omega_{as0}-\omega)$.  $\Omega_e=\sqrt{|\Omega_c|^2-(\gamma_{13}-\gamma_{12})^2}$ is the effective coupling Rabi frequency. $\gamma_e=(\gamma_{12}+\gamma_{13})/2$ is the effective dephasing rate. The third-order nonlinear susceptibility has two resonances separated by $\Omega_e$ and each is associated with a linewidth of $2\gamma_e$. These two resonances here indicate two \keyword{SFWM} paths. In one path the frequency of the anti-Stokes photons is $\omega_{as0}+\Omega_e/2$ and the frequency of the correlated Stokes photons is $\omega_{42}+\Delta_p-\Omega_e/2$. In the other path the anti-Stokes photons have frequency at $\omega_{as0}-\Omega_e/2$ while the paired Stokes photons at $\omega_{42}+\Delta_p+\Omega_e/2$. Consequently, the interference between these two types of \keyword{biphoton}s will appear in the two-photon temporal correlation, as shall be discussed in Sec.~\ref{subsec:RabiOsc}. The results obtained here agree with the dressed-state picture \cite{Wen2,Wen3,Wen4}.

The linear susceptibilities at the anti-Stokes and Stokes frequencies are, respectively,
\begin{eqnarray}
\chi_{as}(\omega_{as0}+\omega)&=&\frac{4N|\mu_{13}|^2(\omega+i\gamma_{12})/(\varepsilon_0\hbar)}{|\Omega_c|^2-4(\omega+i\gamma_{13})(
\omega+i\gamma_{12})},\label{eq:ChiLinearAs} \\
\chi_s(\omega_{s0}-\omega)&=&\frac{N|\mu_{24}|^2(\omega-i\gamma_{13})/(\varepsilon_0\hbar)}{|\Omega_c|^2-4(\omega-i\gamma_{13})(\omega
-i\gamma_{12})}\frac{|\Omega_p|^2}{\Delta_p^2+\gamma_{14}^2},\nonumber \\\label{eq:ChiLinearS}
\end{eqnarray}
where $\Omega_p=\mu_{14}E_p/\hbar$ is the pump Rabi frequency. The complex wave numbers of Stokes and anti-Stokes photons are obtained from the relations $k_s=(\omega_s/c)\sqrt{1+\chi_s}$ and $k_{as}=(\omega_{as}/c)\sqrt{1+\chi_{as}}$, where the imaginary parts indicate the Raman gain and \keyword{EIT} loss, respectively. In an ideal \keyword{EIT} system with zero ground-state dephasing, i.e., with $\gamma_{12}$=0, the linear susceptibility is $\chi_{as}(\omega_{as0})$=0, implying zero linear absorption of the anti-Stokes photons. This allows the nonlinear optics occurring on atomic resonance without absorption and hence enhances the efficiency of the nonlinear interaction.

Taking $|\Omega_p|\ll\Delta_p$, Eqs.~(\ref{eq:ChiLinearAs}) and (\ref{eq:ChiLinearS}) give $k_{as}\simeq{k}_{as0}+\omega/V_g+i\alpha$ and $\chi_s\simeq0$ so that the wave-number mismatching is approximately $\Delta k\simeq\omega/V_g+i\alpha$. Here $k_{as0}$ is the central wave number of the anti-Stokes field, $V_g$ is its group velocity, and $\alpha$, the imaginary part of the anti-Stokes wave number, characterizes the \keyword{EIT} finite loss caused by the non-zero ground-state dephasing rate $\gamma_{12}$. Now Eq.~(\ref{eq:longitudinaldetuningfunction}) can be approximated as
\begin{eqnarray}
\Phi(\omega)\simeq\mathrm{sinc}\Big(\frac{\omega L}{2V_g}+i\frac{\alpha L}{2}\Big)\exp\Big(i\frac{\omega L}{2 V_g}-\frac{\alpha L}{2}\Big).\label{eq:PhaseMatchingApp}
\end{eqnarray}
In Eq.~(\ref{eq:PhaseMatchingApp}), $\alpha=2N\sigma_{13}\gamma_{12}\gamma_{13}/(|\Omega_c|^2+4\gamma_{12}\gamma_{13})$ where $\sigma_{13}=2\pi|\mu_{13}|^2/(\varepsilon_0\hbar\lambda_{13}\gamma_{13})$ is the on-resonance absorption cross section of the transition $|1\rangle\rightarrow|3\rangle$. The longitudinal detuning function of Eq.~(\ref{eq:PhaseMatchingApp}) has a full-width-at-half-maximum (FWHM) phase-matched bandwidth determined by the sinc function, $\Delta\omega_{g}=2\pi\times0.88/\tau_g$, where $\tau_g=L/V_g$ is the anti-Stokes \keyword{group delay} time. The \keyword{group delay} time can be estimated from $\tau_g=L/V_g\simeq(2\gamma_{13}/|\Omega_c|^2)OD$ with the optical depth defined as $OD=N\sigma_{13}L$.

Therefore, there are two important characteristic frequencies that determine the shape of the \keyword{biphoton} waveform. The first is the coupling effective Rabi frequency $\Omega_e$ , which determines the two-resonance spectrum of the nonlinear susceptibility. The second is phase-matching bandwidth $\Delta\omega_{g}$. In the time domain, they correspond to the Rabi time $\tau_r=2\pi/\Omega_e$ and the \keyword{group delay} time $\tau_g$. The competition between $\tau_r$ and $\tau_g$ will determine which effect plays a dominant role in governing the feature of the two-photon correlation. Therefore, in the following we will discuss the two-photon joint-detection measurement in two regimes, damped \keyword{Rabi oscillation} and \keyword{group delay}.

\begin{figure}
\includegraphics[width=0.85\linewidth]{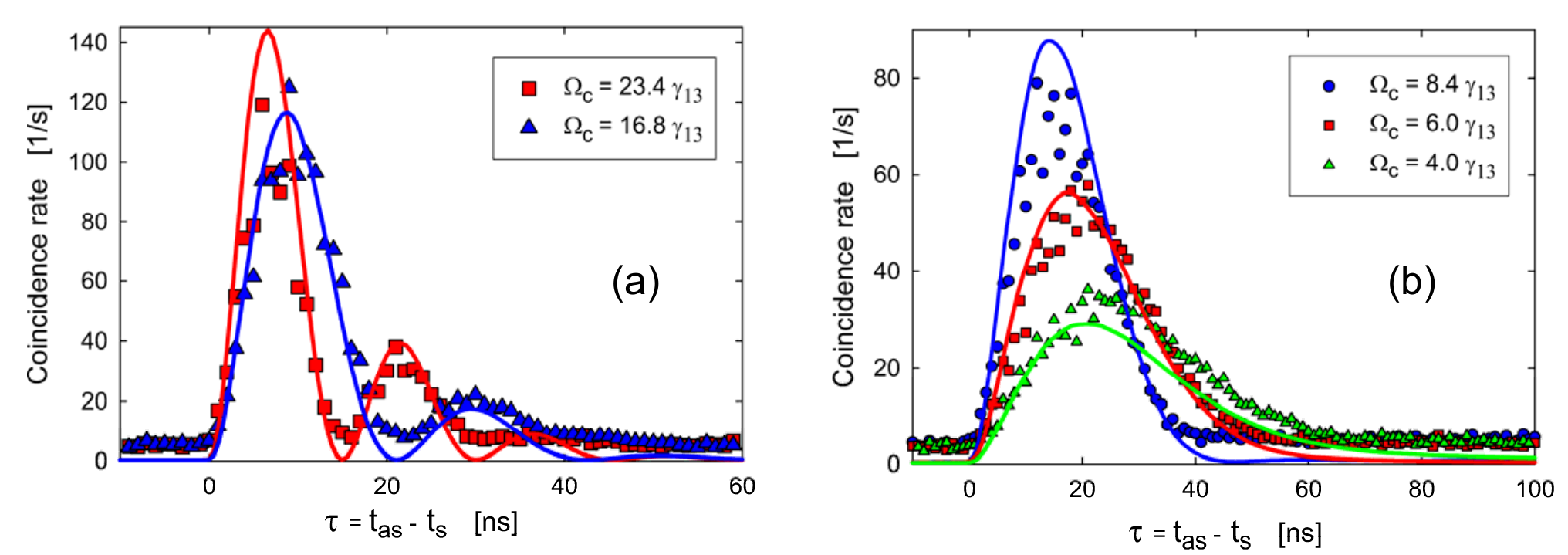}
\centering
\caption{\label{Chuu_fig8}Two-photon correlation function in the damped \keyword{Rabi oscillation} regime. The missing of other oscillation periods in (B) is due to the short dephasing time of about 33 ns. The experimental parameters used here are $\gamma_{13}=\gamma_{14}=2\pi\times3$ MHz, $\gamma_{12}=0.6\gamma_{13}$, $OD=$11, $\Delta_p=-7.5\gamma_{13}$, and $\Omega_p=0.8\gamma_{13}$. Data are taken from Ref.~\cite{BalicPRL2005}.}
\end{figure}

\subsection{\label{subsec:RabiOsc}Damped Rabi Oscillation Regime}

We first look at the regime where the the nonlinear spectrum plays a dominant role and the optical properties of the two-photon amplitude (\ref{eq:BiphotonWaveFunction1}) are mainly determined by the nonlinear coupling coefficient. This regime requires that the effective coupling Rabi frequency $\Omega_e$ be smaller than the phase-matching bandwidth $\Delta\omega_g$, \textit{i.e.},$\Omega_e<\Delta\omega_g$, or equivalently, $\tau_r >\tau_g$, so that we can treat the longitudinal detuning function as $\Phi(\omega)\simeq1$. The \keyword{biphoton} spectral generation rate is proportional to $|\kappa L|^2\propto|\chi^{(3)} L|^2$. Hence both \keyword{biphoton} spectrum intensity and emission rate are proportional to $OD^2$. The two-photon temporal correlation function exhibits a damped \keyword{Rabi oscillation} resulting from the interference between the two resonances of the nonlinear coupling coefficient.

We consider the case of $\Omega_c>|\gamma_{13}-\gamma_{12}|$ first, which implies a real effective coupling Rabi frequency $\Omega_e$. Following Eq.(\ref{eq:BiphotonWaveFunction1}), the two-photon wave amplitude now is determined by the Fourier transform of the nonlinear coupling coefficient (\ref{eq:kappa}) to give
\begin{eqnarray}
\psi(\tau)&=&BL e^{-\gamma_e\tau}e^{-i\varpi_{as}\tau}\sin\Big(\frac{\Omega_e\tau}{2}\Big)\Theta(\tau),\nonumber\\
&=&\frac{i}{2}BL e^{-\gamma_e\tau}e^{-i\varpi_{as}\tau}[e^{-i\Omega_e\tau}-e^{i\Omega_e\tau}]\theta(\tau).
\label{eq:f}
\end{eqnarray}
Here $B=-i\frac{N\mu_{13}\mu_{32}\mu_{24}\mu_{41}\sqrt{\varpi_{as}\varpi_s}}{4c\varepsilon_0\hbar^3\Omega_e(\Delta_p+i
\gamma_{14})}$ and $\Theta(\tau)$ is the Heaviside step function, i.e., $\Theta(\tau)=1$ for $\tau\geq0$, and $\Theta(\tau)=0$ for $\tau<0$. The physics of Eq.~(\ref{eq:f}) is understood as follows. Because the two-photon state is entangled, it cannot be factorized into a function of $t_{as}$ times a function of $t_s$. $|\psi|^2$, depends only on the relative time delay $\tau$, which implies that the pair is randomly generated at any time. The first term in the bracket on the RHS of Eq.~(\ref{eq:f}) represents the two-photon amplitude of paired anti-Stokes at $\omega_{as0}+\Omega_{e}/2$ and Stokes at $\omega_{s0}-\Omega_{e}/2$; while the second term is the two-photon amplitude of paired anti-Stokes at $\omega_{as0}-\Omega_{e}/2$ and Stokes at $\omega_{s0}+\Omega_{e}/2$. Equivalently, this frequency entangled state can be written as $|\omega_{as0}+\Omega_{e}/2\rangle_{as}| \omega_{s0}-\Omega_{e}/2\rangle_{s}-|\omega_{as0}-\Omega_{e}/2\rangle_{as}|\omega_{s0}+\Omega_{e}/2\rangle_{s}$. To further see the interference, let us look at the two-photon Glauber correlation function,
\begin{eqnarray}
G^{(2)}(\tau)=\frac{1}{2}|B L|^2 e^{-2\gamma_e\tau}\big[1-\cos(\Omega_e\tau)\big]\Theta(\tau),\label{eq:RabiG2}
\end{eqnarray}
which displays a damped \keyword{Rabi oscillation} with an oscillation period of $2\pi/\Omega_e$ and a damping rate of $2\gamma_e$. The heaviside function $\Theta(\tau)$ shows that the anti-Stokes photon is always generated after its paired Stokes photon by following the FWM path shown in Fig.~\ref{Chuu_fig7}(a). The two-photon correlation function also shows the well-known anti-bunching effect [$G^{(2)}(0)\leq G^{(2)}(\tau)$]. Similarly to the polarization entangled Bell states, the visibility of the \keyword{Rabi oscillation}, resulting from the two-photon interference in time domain, can be taken as an evidence for the time-frequency entanglement.

Figure~\ref{Chuu_fig8}(a) shows the first experimental demonstration of this type of \keyword{biphoton} source by Bali\'{c} \textit{et al.} \cite{BalicPRL2005}. Figure~\ref{Chuu_fig8}(a) corresponds to the case of $\tau_e>\tau_r$ and the oscillations are clearly resolved. Figure ~\ref{Chuu_fig8}(b) is the case of $\tau_e<\tau_r$ where only the first oscillation is observable and other disappear because of the fast dephasing rate $\gamma_e$.

\keyword{Rabi oscillation}s have also been observed in three-level and two-level atomic systems \cite{HarrisPRL2006, DuPRL2011Entanglement, DuPRL2007}.
\\

\subsection{Group Delay Regime}

Suggested by Balic \textit{et al.}\cite{BalicPRL2005} and demonstrated by Du \textit{et al.}\cite{Subnatural}, the \keyword{group delay} regime is defined as $\tau_g>\tau_r$ and the \keyword{EIT} slow-light effect can be used to control the \keyword{biphoton} temporal coherence time. The \keyword{group delay} condition is equivalent to $\Omega_e>\Delta\omega_g\approx2\pi/\tau_g$, i.e., the \keyword{biphoton} bandwidth is determined by phase matching. For this reason, in this subsection we treat the third-order nonlinear susceptibility as a constant over the phase-matching spectrum. As a consequence, the double resonances of \keyword{biphoton} generation are suppressed. The two-photon correlation tends to be rectangular shaped, more like that of the conventional SPDC photons. We obtain the \keyword{biphoton} wave function approximately as $\psi(\tau)\simeq\kappa_0L\tilde{\Phi}(\tau)$, where $\kappa_0$ is the on-resonance nonlinear coupling constant and $\tilde{\Phi}(\tau)=1/(2\pi)\int\Phi(\omega)e^{-i\omega\tau}d\omega$ is Fourier transform of the longitudinal detuning function. When the \keyword{EIT} bandwidth [$\Delta\omega_{tr}\simeq|\Omega_c|^2/(2\gamma_{13}\sqrt{OD})$] is wider than the phase-matching bandwidth, the anti-Stokes loss is negligible and the two-photon wave function approaches a rectangular shape. However, when the \keyword{EIT} loss is significant, the \keyword{biphoton} waveform follows an exponential decay.

\begin{figure}
\includegraphics[width=0.55\linewidth]{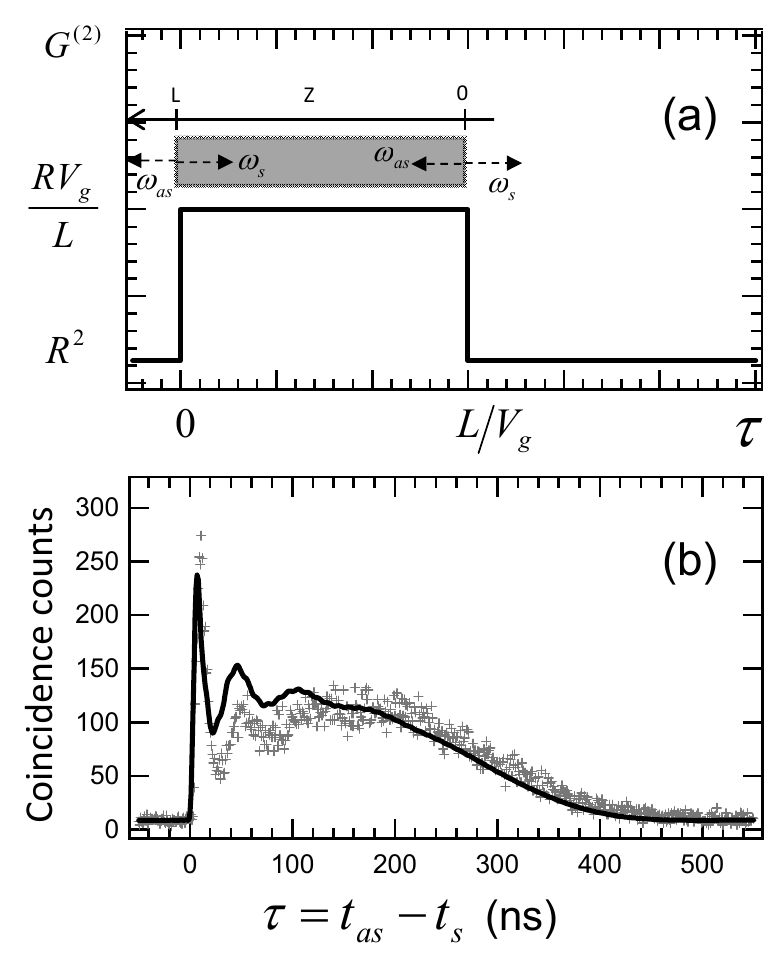}
\centering
\caption{\label{Chuu_fig9}Two-photon correlation function in the \keyword{group delay} regime. (a) Ideal rectangular-shape correlation with a relative \keyword{group delay} $L/V_g$ and pair emission rate $R$. (b) Experimental data (red) and theoretical curve (solid blue line) obtained from the Heisenberg-Langevin theory. Data are taken from Ref.~\cite{Subnatural}.}
\end{figure}

When $\Delta\omega_{tr}>\Delta\omega_g$, we ignore the EIT linear loss and rewrite  Eq.~(\ref{eq:PhaseMatchingApp}) as
\begin{eqnarray}
\Phi(\omega)\simeq\mathrm{sinc}\Big(\frac{\omega L}{2V_g}\Big)\exp\Big(i\frac{\omega L}{2V_g}\Big).
 \label{eq:PhaseMatchingGDR1}
\end{eqnarray}
We then obtain the \keyword{biphoton} wave function
\begin{eqnarray}
\psi(\tau)\simeq\kappa_{0}L\tilde{\Phi}(\tau)=\kappa_{0}V_g \Pi(\tau;0,L/V_g)e^{-i\varpi_{as}\tau}.
\label{eq:SquareFunc}
\end{eqnarray}
The rectangular function is defined as $\Pi(\tau;\tau_1,\tau_2)=\theta(\tau-\tau_1)-\theta(\tau-\tau_2)$. Equation (\ref{eq:SquareFunc}) shows that the anti-Stokes photon is always delayed with respect to its paired Stokes photon because of the slow light effect. The two-photon correlation time is determined by the \keyword{group delay} $\tau_g=L/V_g$. Using Eq.~(\ref{eq:PhotonPairrate}) we get the photon pair generation rate
\begin{eqnarray}
R=|\kappa_0|^2V_gL.\label{eq:PairRateGDR}
\end{eqnarray}
Thus in the group delay regime the total rate of paired counts scales linearly as $OD$ even though the spectral generation rate, $\kappa_0L\Phi(\omega)$, scales as $OD^2$. This is because the bandwidth reduces linearly with optical depth \cite{BalicPRL2005}. As illustrated in Fig.~\ref{Chuu_fig9}(a), the rectangular-shape waveform can be understood in the following picture. When the photon pair is produced from the front surface ($z=L$), the anti-Stokes photon has no delay and both photons arrive at detectors simultaneously. When emitted from the back surface ($z=0$), the anti-Stokes photon is delayed relative to the Stokes photon by $\tau_g$. Since the photon pair generation probability density is evenly distributed in the medium, a rectangular shape shows up in the biphoton temporal waveform. Figure~\ref{Chuu_fig9}(a) shows an ideal rectangular-shape correlation with a \keyword{group delay} of $\tau_g=L/V_g$. For conventional SPDC photons, the sub-ps rectangular-shaped \keyword{biphoton} waveform has only been indirectly confirmed in \cite{Sergienko}. For the SFWM in cold atoms discussed here, we find that to observe the rectangular shape, the condition $\Delta\omega_{tr}>\Delta\omega_g$ sets a lower bound for the optical depth. Using $\Delta\omega_{tr}\simeq|\Omega_c|^2/(2\gamma_{13}\sqrt{OD})$ and $\Delta\omega_g\simeq2\pi/\tau_g\simeq\pi|\Omega_c|^2/(\gamma_{13}OD)$, we obtain $OD>4\pi^2$. Or equivalently, it requires the \keyword{EIT} delay-bandwidth product greater than two \cite{NLOHarris}.

\begin{figure}
\includegraphics[width=0.55\linewidth]{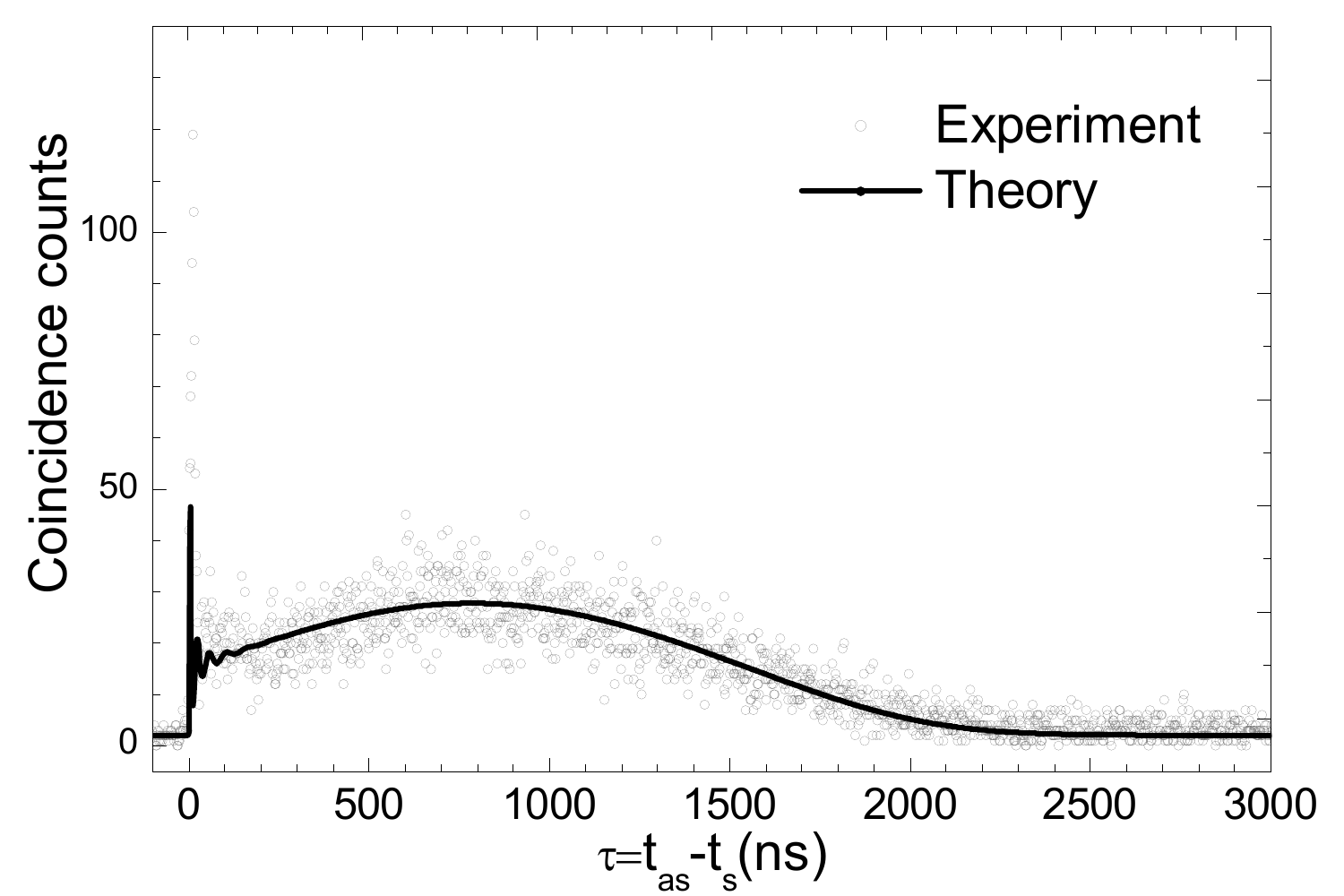}
\centering
\caption{\label{Chuu_fig10} Biphotons with coherence time up to about 2 $\mu$s. Operating parameters are: OD=130, $\Omega_{c}=2\pi\times7.77$ MHz and $\Omega_{p}=2\pi\times1.14$ MHz. Data are taken from Ref. \cite{DuOptica2014}.}
\end{figure}

In Fig.~\ref{Chuu_fig9}(b) the experimental demonstration of a near-rectangular-shape correlation is obtained at the optical depth around 53 \cite{Subnatural}. The experimental parameters used here are $\gamma_{13}=\gamma_{14}=2\pi\times3$ MHz, $\gamma_{12}=0.02\gamma_{13}$, $OD=53$, $\Delta_p=48.67\gamma_{13}$, $\Omega_p=1.16\gamma_{13}$, and $\Omega_c=4.20\gamma_{13}$, respectively. The \keyword{EIT} transparency width is estimated around $3.63$ MHz and the phase-matching spectral width about $2.93$ MHz. It is found that the exponential-decay behavior at the tail is due to the finite \keyword{EIT} loss, which alters the correlation function away from the ideal rectangular shape.

The sharp peak shown in the leading edge of the two-photon coincidence counts in Fig.~\ref{Chuu_fig9}(b) is the first observed Sommerfeld-Brillouin \keyword{precursor} at the \keyword{biphoton} level \cite{BiphotonPrecursor}. In the stationary-phase approximation, starting from Eq.~(\ref{eq:BiphotonWaveFunction1}) one can show that the sharp peak results from the beating of \keyword{biphoton}s which are generated outside of the \keyword{EIT} opacity window. The detailed analysis of the \keyword{precursor} generation at the two-photon level has been presented in \cite{BiphotonPrecursor}.

Following Eq.~(\ref{eq:SquareFunc}), one expects the length $L/V_g$ of the rectangular function increases if one reduces the anti-Stokes group velocity $V_g$ by reducing the coupling laser power. However, Eq.~(\ref{eq:SquareFunc}) is valid only when the \keyword{EIT} is preserved. When the \keyword{EIT} loss becomes significant, the \keyword{biphoton} has an exponential decay waveform \cite{DuJOSAB2008}. Therefore at a certain OD, there is a limitation to prolonging the coherence time while maintaining the rectangular-like shape. Most recently, by pushing the OD to 130, the two-photon coherence time was extended to nearly 2 $\mu$s, as shown in Fig.~\ref{Chuu_fig10}, which corresponds to a FWHM bandwidth of 0.43 MHz \cite{DuOptica2014}.

\section{\label{subsec:Manipulation}Manipulation of Narrowband Single Photons}

The long coherence time of these narrow-band \keyword{biphoton}s allows us to not only directly resolve their temporal waveform with existing commercial single photon detectors (with a typical resolution of 1 ns) but also manipulate their quantum waveform with external phase-amplitude modulators \cite{EOMSinglephoton, BelthangadyPRA2009}. In addition, compared with SPDC using a single pump laser, there are more freedoms to manipulate the the \keyword{biphoton} generation in \keyword{SFWM} with two driving lasers. We will show in this section how to engineer the \keyword{biphoton} quantum states by controlling the temporal \cite{DuPRA2009, DuPRL2010Biphoton} and spatial patterns \cite{DuOptica2014, DuShapingBiphoton2014} as well as the polarizations \cite{DuPRL2011Entanglement, YanPRL2014} of the classical driving fields.\\

\noindent\textbf{1) Electro-Optical Modulation of Heralded Single Photons}\\

\begin{figure}[htbp]
\includegraphics [width=0.85\linewidth]{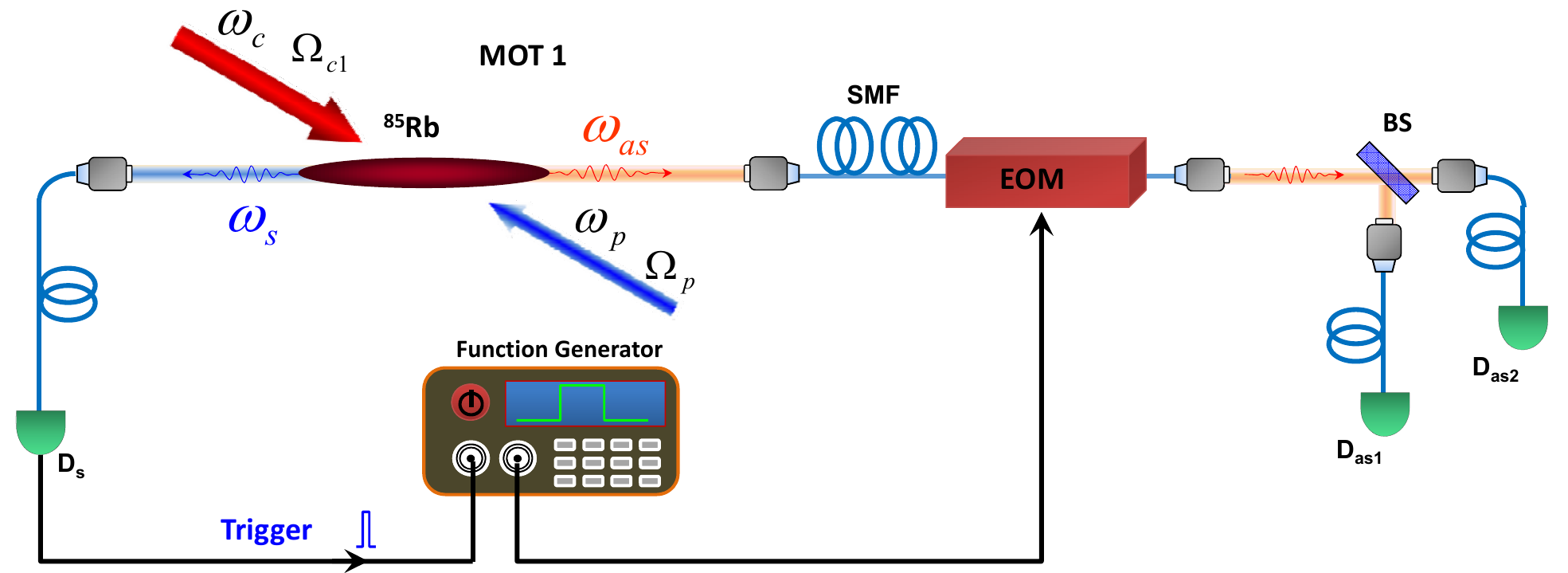}
\centering
\caption{Schematic of heralded single photon generation and conditional modulation with an EOM.} \label{Chuu_fig11}
\end{figure}

\begin{figure}[htbp]
\includegraphics[width=0.55\linewidth]{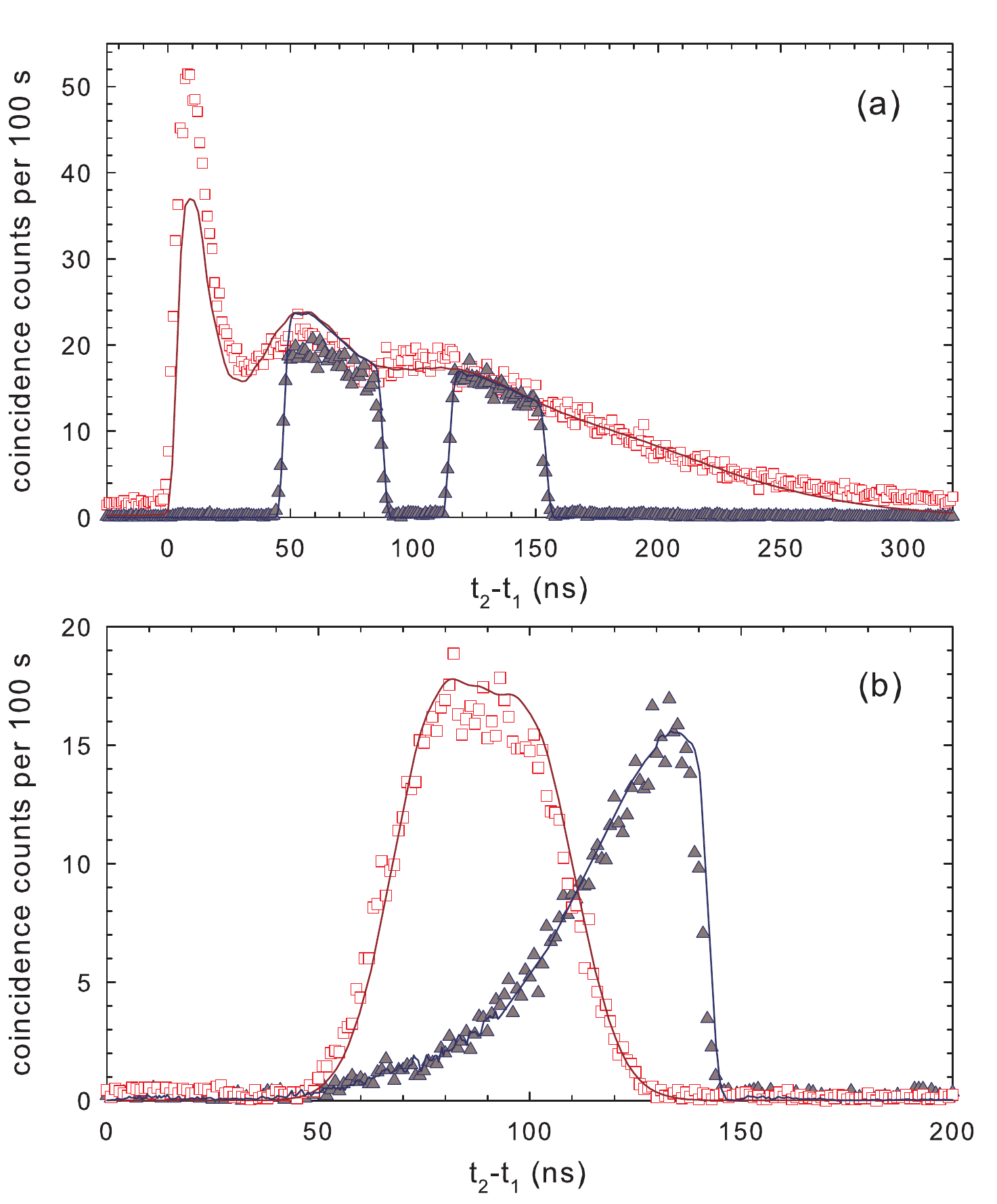}
\centering
\caption{Waveforms of heralded anti-Stokes photons. (a) Modulated and unmodulated waveforms. (b)  Waveforms with Gaussian and rising exponential shapes. Data are taken from \cite{EOMSinglephoton}.} \label{Chuu_fig12}
\end{figure}

It is well know that time-frequency entangled photon pairs can be used to efficiently produce heralded single photons with well defined relative time origin. Here we describe how single photons may be modulated so as to produce single-photon waveforms whose amplitude and phase are functions of time. As shown in Fig.~\ref{Chuu_fig11}, The detection of a Stokes photon at $D_{1}$ triggers the function generator that drives the electro-optic modulator (EOM) to shape the waveform of the anti-Stokes photon. In this way the heralded anti-Stokes photon temporal wave function (phase and amplitude) may be shaped in the same manner as one modulates a classical light pulse.

When a Stokes photon is detected at D$_s$ ($z_{s}=0$), the heralded single anti-Stokes photon wave packet shaped by the EOAM can be described as
\begin{eqnarray}
\label{eq:HeraldedWavePacket}
\Psi(z,\tau)&=&\langle0|m(\tau)\hat{a}_{as}(z,\tau)\hat{a}_{s}(0,0)|\Psi_{s,as}\rangle \nonumber \\
&=&m(\tau-z/c)\psi(\tau-z/c)e^{i(k_{as0}z-\omega_{as0} \tau)}.
\end{eqnarray}
where $|\Psi_{s,as}\rangle$ is the two-photon time-frequency entangled state \cite{DuJOSAB2008}, and $m(\tau)$ is the modulation. It is clear that the heralded single-photon waveform is directly modulated by the EOM. Therefore, within the coherence time, the single-photon waveform can be arbitrarily shaped. As shown in Fig.~\ref{Chuu_fig12}, The heralded single photons can be modulated into two rectangular pulses, Gaussian or time reversed exponential \cite{EOMSinglephoton}.

The importance of the electro-optic method is its speed and ability to modulate phase as well as amplitude. The technique provides the technology for studying the response of atoms to shaped single photon waveforms on a time-scale comparable to the natural linewidth.\\

\noindent\textbf{2) Quantum Fourier Cosine Transform: Modulation and Measurement of Biphoton Waveform}\\

\begin{figure}[h]
\includegraphics[width=0.85\linewidth]{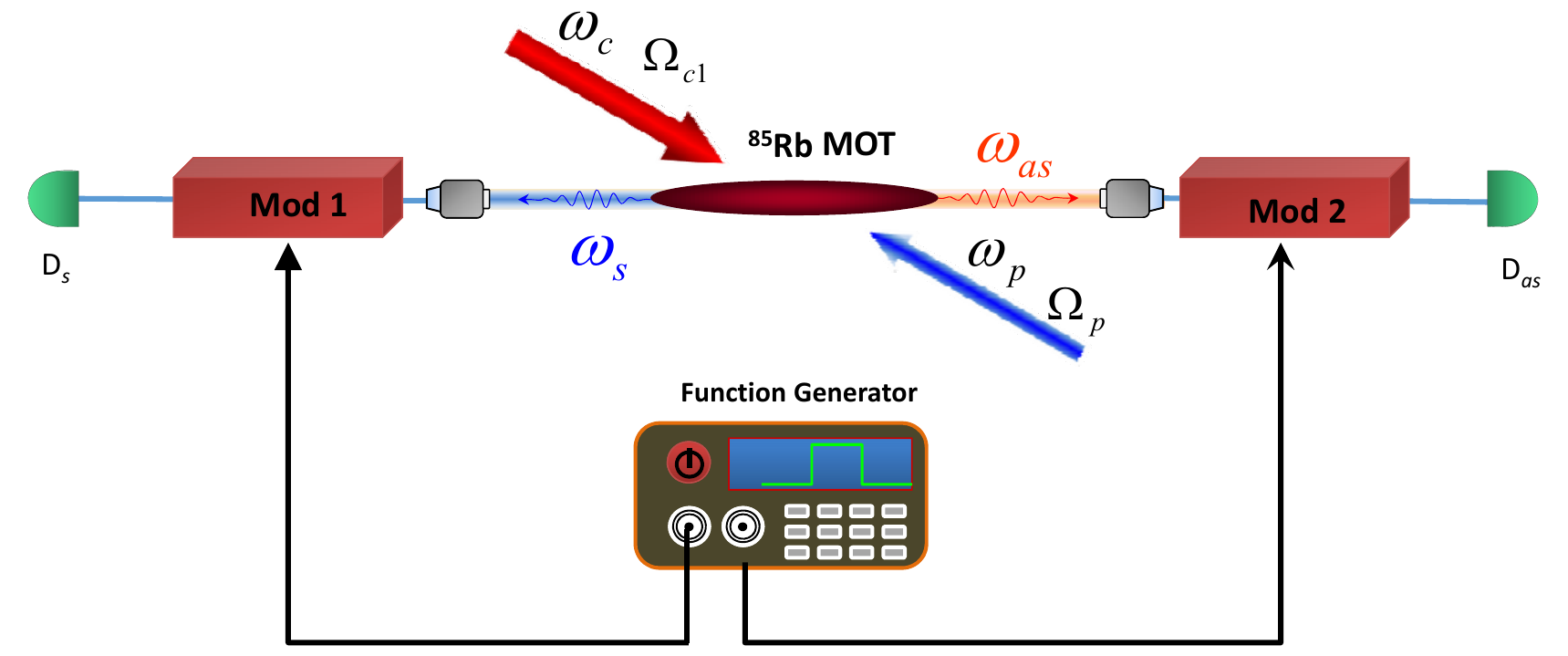}
\centering
\caption{\label{Chuu_fig13} Modulation and measurement of time-energy entangled photons synchronously driven sinusoidal modulators and slow detectors.}
\end{figure}

As described previously, the Stokes-anti-Stokes two-photon temporal waveform is directly measured as coincidence counts between two single photon detectors. However, if the speed of the detectors is slow as compared to the two-photon correlation time, the \keyword{biphoton} temporal waveform can not resolved by the resolution of the detectors. That is the reason why the wide-band SPDC \keyword{biphoton} waveform can not be directly measured by the existing commercial single photon detectors. Here we describe an approach to the problem of measurement of \keyword{biphoton} waveforms using slow detectors. Figure \ref{Chuu_fig13} shows the essential idea. The Stokes and anti-Stokes photons are incident on synchronously driven sinusoidal amplitude modulators (MOD1 and MOD2). The coincidence count rate between single-photon counting modules (SPCMs) is measured as a function of the sinusoidal modulation frequency. The SPCMs are slow as compared to the pulse width of the \keyword{biphoton} waveform. With $\tau$ equal to the relative arrival time of the Stokes and anti-Stokes photons, the inverse Fourier transform of the (frequency domain) measurement of coincidence counts versus frequency yields the Glauber correlation function $G^{(2)}(\tau)$ and therefore the square of the absolute value of the \keyword{biphoton} wave function \cite{BelthangadyPRA2009}.

The modulated correlation function can be written as
\begin{eqnarray}
\label{eq:MCF}
G^{(2)}_M(t,t+\tau)=|m_1(t)|^2|m_2(t+\tau)|^2G^{(2)}_0(\tau),
\end{eqnarray}
where $m_1$ and $m_2$ are the two amplitude modulation functions. $G^{(2)}_0(\tau)$ is the unmodulated Glauber correlation function. We average $G^{(2)}_M(t,t+\tau)$ over a period $T$ of the modulating frequency to form the time averaged correlation function
\begin{eqnarray}
\label{eq:AveragedMCF}
\overline{G^{(2)}_M(\tau)}=\frac{1}{T}\int_0^T G^{(2)}_M(t,t_\tau)dt.
\end{eqnarray}
Combining Eqs.(\ref{eq:MCF}) and (\ref{eq:AveragedMCF}), we obtain
\begin{eqnarray}
\label{eq:AveragedMCF1}
\overline{G^{(2)}_M(\tau)}=M(\tau)G^{(2)}_0(\tau),\nonumber\\
M(\tau)=\frac{1}{T}\int_0^T |m_1(t)|^2|m_2(t+\tau)|^2dt
\end{eqnarray}
where $M(\tau)$ is the intensity correlation function of the modulators. Here we assume both channels are modulated by the same sinusoidal amplitude modulation $m_1(t)=m_2(t)=\cos(\omega t+\varphi)$. Then we have $M(\tau)=1/4+1/8\cos(2\omega\tau)$, and the integrated coincidence count becomes
\begin{eqnarray}
\label{eq:IntegratedMCF}
\int_0^\infty \overline{G^{(2)}_M(\tau)} d\tau=\frac{1}{8}\int_0^\infty [2+\cos(2\omega\tau)]G^{(0)}(\tau) d\tau.
\end{eqnarray}
We neglect the dc term and normalize to obtain the Fourier cosine transform pair
\begin{eqnarray}
\label{eq:FourierCosineT}
F(2\omega)=\sqrt{\frac{2}{\pi}}\int_0^\infty G^{(0)}(\tau)\cos(2\omega\tau)d\tau,\nonumber\\
G^{(0)}(\tau)=\sqrt{\frac{2}{\pi}}\int_0^\infty F(2\omega)\cos(2\omega\tau)d\omega.
\end{eqnarray}
Therefore the two-photon correlation function can be obtained by an inverse Fourier cosine transform from integrated coincidence counts (by slow detectors).

The proof of principle of the above Fourier technique in measuring \keyword{biphoton} waveform was demonstrated with narrow-band \keyword{biphoton}s and the slow detectors are simulated with integration \cite{BelthangadyPRA2009}. The result agrees well with the directly measured waveforms. For practical applications, this technique can be extended to short \keyword{biphoton}s.\\

\noindent\textbf{3) Shaping Biphoton Temporal Waveforms with Temporally Modulated Classical Fields}\\
\begin{figure}[h]
\includegraphics[width=0.85\linewidth]{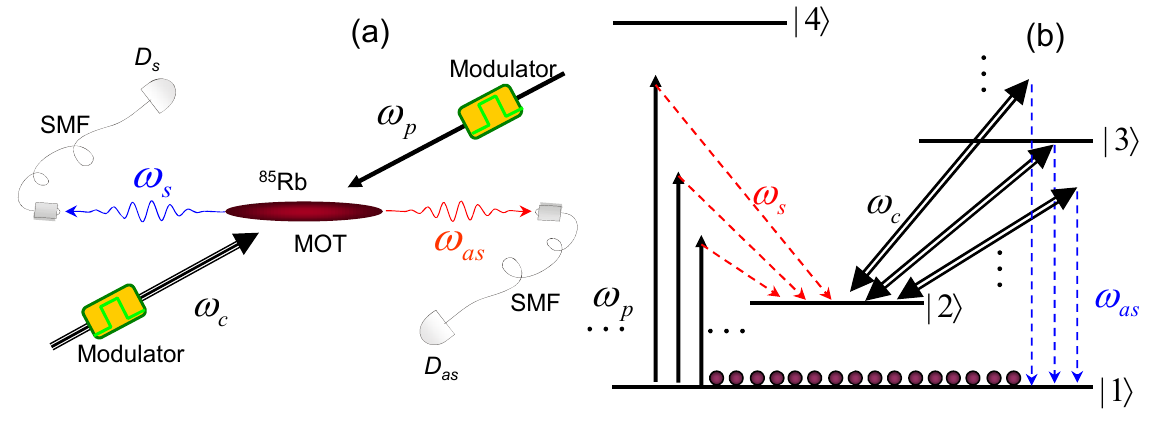}
\centering
\caption{\label{Chuu_fig14}Schematic of \keyword{biphoton} generation with modulated driving-laser fields. (a) Experimental configuration. (b) Atomic energy level diagram for multichannel \keyword{biphoton} generation.}
\end{figure}

By passing long single photons through electro-optical modulators (EOM), it is possible to modulate one of the paired photons or their correlation function. However, these external modulators always introduce losses and attenuations. As shown in Fig.~\ref{Chuu_fig7}, due to the two-photon resonance condition in the \keyword{SFWM} process in the double-$\Lambda$ atomic system, the central frequency of the Stokes photon follows the pump laser, while the central frequency of the anti-Stokes photon follows the coupling laser (in the group delay regime). As a result, if we add a modulation to the pump (coupling) laser, the modulation will affect the generated Stokes (anti-Stokes) photon. As compared to the SPDC source with a single pump laser, the \keyword{SFWM} scheme with two driving lasers allows more freedoms to manipulate the \keyword{biphoton} state by controlling the two classical driving lasers.

When the pump and coupling lasers are periodically modulated, As shown in Fig.~\ref{Chuu_fig14}(b), the driving laser fields are decomposed in frequency domain into discrete frequency components. As a result, \keyword{biphoton} generation follows many possible \keyword{SFWM} paths. The interference between these multichannel FWMs provides a controllable way to manipulate and engineer the \keyword{biphoton} wave packets. Under the condition that the atomic optical depth in the transition $|1\rangle\rightarrow|3\rangle$ is high, the \keyword{EIT} slow light effect is significant, and there is no spectral overlap between these \keyword{SFWM} channels, following the theory by Du \textit{et al.} \cite{DuPRA2009}, the \keyword{biphoton} wave amplitude can be obtained analytically
\begin{eqnarray}
\Psi(t_s,t_{as})=\frac{c\varepsilon_0}{2\sqrt{\overline{I}_p\overline{I}_c}} E_p(t_s)E_c(t_{as})\psi_0(t_{as}-t_s)\times e^{-i\omega_{s0}t_s}e^{-i\omega_{as0}t_{as}},
\label{eq:BiphotonWaveFunctionM}
\end{eqnarray}
where $E_p(t_s)$ and $E_c(t_{as})$ are the complex envelopes of the pump and coupling laser fields. $\overline{I}_p$ and $\overline{I}_c$ are the pump and coupling laser average intensities. $\psi_0(t_{as}-t_s)e^{-i\omega_{s0}t_s}e^{-i\omega_{as0}t_as}$ is the original two-photon wave packet without modulation on the driving fields. As shown in Eq.~(\ref{eq:BiphotonWaveFunctionM}), the two-photon time-frequency entanglement information is preserved while the pump and coupling field profiles are mapped into the \keyword{biphoton} waveform. The Stokes and anti-Stokes temporal correlation function is
\begin{eqnarray}
G^{(2)}(t_s,t_{as})\equiv|\Psi(t_s,t_{as})|^2=\frac{I_p(t_s)I_c(t_{as})}{\overline{I}_p\overline{I}_c}G^{(2)}_0(t_{as}-t_s), \nonumber\\
\label{eq:Correlation}
\end{eqnarray}
where $I_p(t_s)$ and $I_c(t_{as})$ are the pump and coupling laser intensity temporal profiles. $G^{(2)}_0(\tau)=|\psi_0(\tau)|^2$ is the correlation function without modulation. Therefore the time-averaged correlation becomes
\begin{eqnarray}
\overline{R}(\tau)=\mathrm{C}(\tau)R_0(\tau),
\label{eq:CorrelationAveraged}
\end{eqnarray}
where $\mathrm{C}(\tau)\equiv\lim_{\Delta T\rightarrow\infty}\frac{1}{\Delta T \overline{I}_p\overline{I}_c}\int_0^{\Delta T}I_p(t)I_c(t+\tau)dt$ is the time-averaged pump-coupling correlation function. Equations (\ref{eq:BiphotonWaveFunctionM})-(\ref{eq:CorrelationAveraged}) show that it is possible to manipulate the \keyword{biphoton} temporal wave packet and its correlation functions in a controllable way. The proof of principle has been experimentally demonstrated in Ref.~\cite{DuPRL2010Biphoton}.\\

\noindent\textbf{4) Shaping Biphoton Temporal Waveforms with Spatially Modulated Classical Fields}\\

Following Rubin's \keyword{group delay} picture \cite{RubinPRA1994}, in the ideal \keyword{group delay} regime, the rectangular waveform reflects the uniform spatial distribution of the photon pair generation probability. This is indeed ensured in our previous theoretical model where the atomic density, pump and coupling laser fields are assumed to be uniform. It is noticed that the \keyword{biphoton} generation probability is proportional to the pump field intensity. Therefore, if the pump field is spatially modulated along the longitudinal direction of the \keyword{biphoton} generation, this spatial modulation will affect the \keyword{biphoton} temporal waveform. This effect was first reported in a recent experiment for achieving the two-photon coherence time up to about 2 $\mu$s \cite{DuOptica2014}. In this experiment, the transverse Gaussian amplitude profile of the pump laser beam is projected to the \keyword{biphoton} longitudinal direction and revealed in the two-photon correlation function. This effect becomes only apparent at a high OD of more than 100 -- at OD$<$60, the theory in Sec. \ref{sec:General Formulism} still holds and the pump field can be treated as a plane wave even though in reality it is in a gaussian mode.

\begin{figure}[h]
\centering
\includegraphics[width=0.55\linewidth]{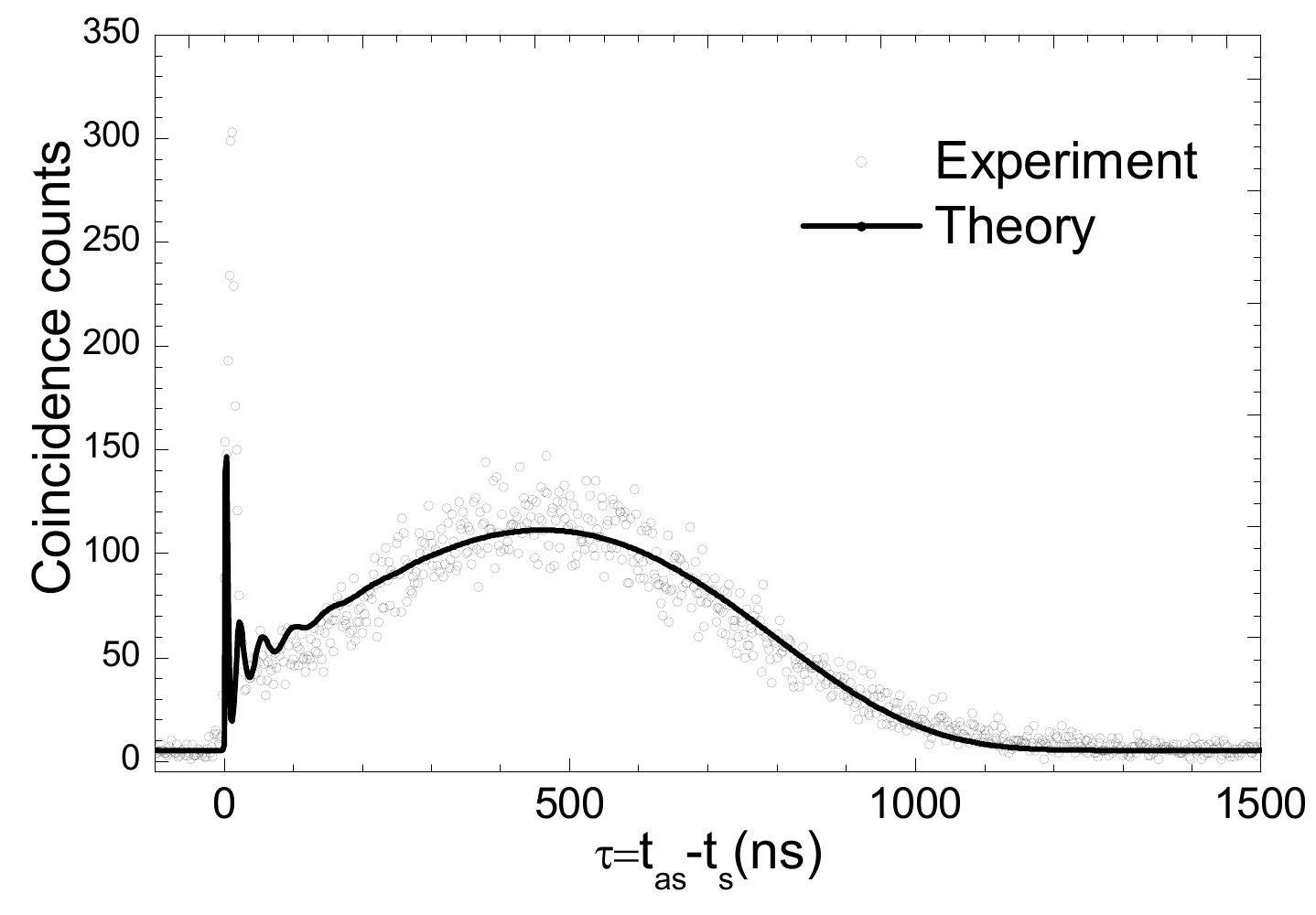}
\centering
\caption{\label{Chuu_fig15} Biphoton temporal waveform reveals the Spatial Gaussian profile of the pump laser beam. Operating parameters are: OD=130, $\Omega_{c}=2\pi\times11.34$ MHz and $\Omega_{p}=2\pi\times1.14$ MHz. Data are taken from Ref. \cite{DuOptica2014}}
\end{figure}

In Ref.~\cite{DuOptica2014}, the pump-coupling laser beams are aligned with am angle of $2.8^o$ to the \keyword{biphoton} longitudinal z-axis and the pump beam transverse Gaussian profile is projected to the z-axis. Taking into account the pump field profile effect, we modify Eq.~(\ref{eq:BiphotonWaveFunction}) to have
\begin{eqnarray}
\psi(\tau)=\frac{1}{2\pi}\int d\omega \kappa(\omega)\mathrm{F}(\Delta k)e^{i(k_{as}+k_s)L/2}e^{-i\omega\tau},
 \label{eq:BiphotonWaveFunctionF}
\end{eqnarray}
where the longitudinal detuning function is replaced by $\mathrm{F}(\Delta k)e^{i(k_{as}+k_s)L/2}$. $\mathrm{F}(\Delta k)$ is the Fourier transform of the pump field profile $f(z)=1/(2\pi)\int dk F(k)e^{ikz}$ along the z-axis. In the \keyword{group delay} regime, the spatial phase propagation in Eq. (\ref{eq:BiphotonWaveFunctionF}) can approximated as
\begin{eqnarray}
(k_{as}+k_s)L/2\simeq \phi_0+\omega \tau_g/2,
 \label{eq:Phase}
\end{eqnarray}
where $\phi_0$ is a constant phase factor. The two-photon spectrum is mainly determined by the phase-matching longitudinal function $\mathrm{F}[\Delta k(\omega)]$, and $\kappa(\omega)\simeq\kappa_0$ varies slowly in frequency. The we can reduce Eq.(\ref{eq:BiphotonWaveFunctionF}) to
\begin{eqnarray}
\psi(\tau)\simeq\kappa_0 V_g f(L/2-V_g\tau)e^{i\phi_0}.
 \label{eq:BiphotonWaveFunctionF2}
\end{eqnarray}
It is clear that the pump field spatial variation is mapped into the two-photon quantum temporal waveform with its origin delayed by $L/(2V_g)=\tau_g/2$. The two-photon temporal correlation time is determined by the \keyword{group delay} $\tau_g$ of the slow anti-Stokes photon.

Figure \ref{Chuu_fig15} shows the experimental results at OD=130 \cite{DuOptica2014}. There are two main features of the two-photon correlation function. The fast oscillating spike at the leading edge is the \keyword{biphoton} optical \keyword{precursor} which travels at the speed of light in vacuum \cite{BiphotonPrecursor}. The later, slowly varying long waveform is generated from the narrow \keyword{EIT} window. The Gaussian shape reveals the pump laser intensity profile as we expected from Eq.~(\ref{eq:BiphotonWaveFunctionF2}).

The detailed discussion of the effects of the nonuniformity of the driving fields on the \keyword{biphoton} waveform and \keyword{biphoton} engineering with spatially modulated driving lasers can be found in \cite{DuShapingBiphoton2014}.\\

\noindent\textbf{5) Polarization Entanglement}\\

The \keyword{SFWM} scheme provides a natural entanglement mechanism in the time-frequency domain, but it is extremely difficult to produce polarization entanglement because of the polarization selectivity of \keyword{EIT} in a non-polarized atomic medium \cite{YuPRA2000}. The ``writing-reading'' technique with optical pumping provides a solution to polarization entanglement but results in reducing time-frequency entanglement \cite{KuzmichScience2004}.

\begin{figure}[h]
\includegraphics[width=0.9\linewidth]{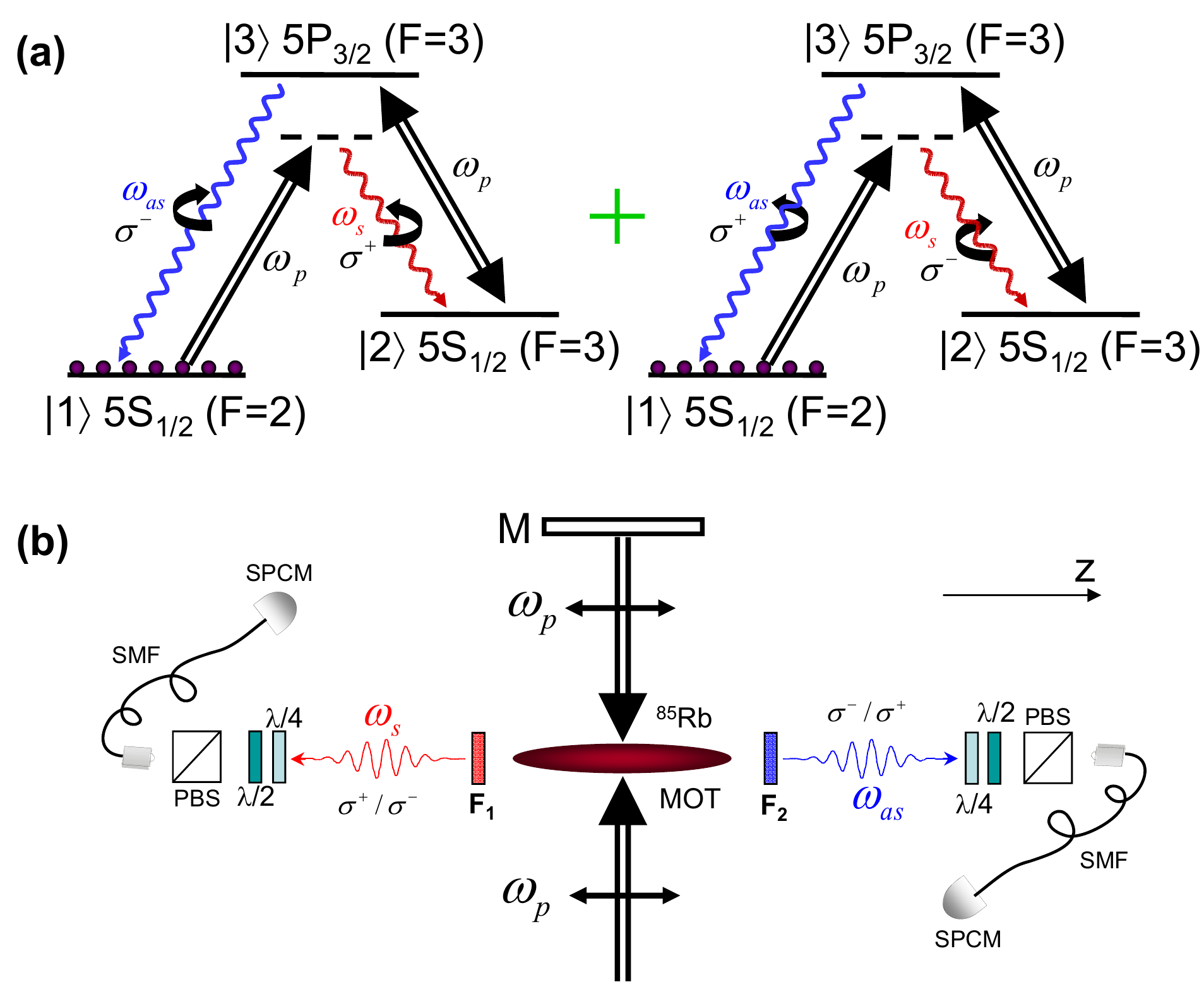}
\centering
\caption{\label{Chuu_fig16}Polarization entanglement generation in a right-angle \keyword{SFWM} configuration. (a)$^{85}$Rb energy level diagram with two possible polarization configurations for the spontaneously emitted photon pairs. (b) Experimental setup with a right-angle geometry. Two sets of quarter-wave plates, half-wave plates, and PBSs are inserted for measuring polarization correlation and quantum state tomography. The figure is taken from Ref. \cite{DuPRL2011Entanglement}.}
\end{figure}

\begin{figure}
\includegraphics[width=\linewidth]{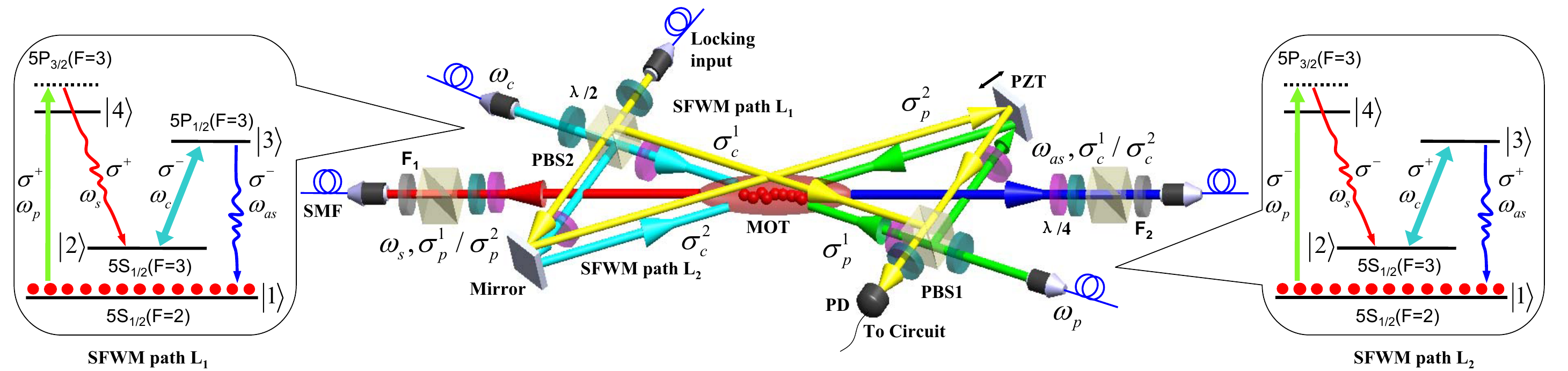}
\centering
\caption{\label{Chuu_fig17}Experimental setup for
producing subnatural-linewidth polarization-entangled photon pairs. The polarization entanglement is
created by the quantum interference of the two spatially symmetric
\keyword{SFWM} processes driven by two counter-propagating pump-coupling beams
($L_1$ and $L_2$). The inserted energy level
diagrams are two possible \keyword{SFWM} channels for $L_{1}$ and $L_{2}$,
respectively. The figure is taken from Ref. \cite{YanPRL2014}.}
\end{figure}

Figure ~\ref{Chuu_fig16} shows a schematics for polarization entanglement generation in a right angle geometry \cite{DuPRL2011Entanglement}. A single pump laser is retro-reflected and serve both pump and coupling laser beams in the \keyword{SFWM} process. The perfect \keyword{phase matching} condition allows spontaneously generated paired photons to be emitted at right angle. To produce entanglement in polarization, we make use of the degenerate Zeeman sub states of each hyperfine energy level. By choosing the 2D MOT longitudinal symmetry axis as the quantization axis (z-axis), Zeeman states with $\Delta M_F=0$ are coupled by the linearly-polarized pump beams. Conservation of angular moment along the z-axis allows two possible circular polarization configurations as shown in Fig.~\ref{Chuu_fig16}(a): $|\sigma_{s}^{+}\sigma_{as}^{-}\rangle$ and $|\sigma_{s}^{-}\sigma_{as}^{+}\rangle$. From the symmetry of our system, the quantum state of the paired Stokes and anti-Stokes photons at the two detectors are described by
\begin{eqnarray}
|\Psi_{s,as}(t_s,t_{as})\rangle=\psi(\tau)e^{-i\omega_{s0}t_s}e^{-i\omega_{as0}t_{as}}\times\frac{1}{\sqrt{2}}(|\sigma^+_s\sigma^-_{as}\rangle+|\sigma^-_s\sigma^+_{as}\rangle).\label{eq:state}
\end{eqnarray}
However, in this right angle configuration, because of some forbidden transition between the Zeeman sub states which can not be coupled by the coupling laser, the anti-Stokes photons do not see a complete \keyword{EIT}. As a result, the system can only work at a low OD in the \keyword{Rabi oscillation} regime \cite{DuPRL2011Entanglement}.

Figure ~\ref{Chuu_fig17} shows a more robust scheme for generating polarization entanglement for narrowband \keyword{biphoton}s in the \keyword{group delay} regime \cite{YanPRL2014}. The pump laser beam is equally split into two beams after the first polarization beam splitter (PBS1). These two beams, with opposite circular polarizations ($\sigma ^ + $ and $\sigma ^ -$) after two quarter-wave plates, then intersect at the MOT with an angle of $\pm 2.5^\circ$  to the longitudinal axis. Similarly, the two
coupling laser beams after PBS2 with opposite circular polarizations overlap with the two pump beams from opposite directions. In presence of these two pairs of counter-propagating
pump-coupling beams, phase-matched Stokes and anti-Stokes paired photons are produced along the
longitudinal axis. In each \keyword{SFWM} path, the polarizations of the Stokes and anti-Stokes photons match those of the corresponding pump and coupling fields.

To obtain the polarization entanglement, we must stabilize the phase difference between the two \keyword{SFWM} spatial paths. This is achieved by injecting a reference laser beam from the second input of PBS2. The two reference beams split after PBS2 are then recombined after PBS1 and detected by a photo detector (PD, a half wave plate and a PBS are used to obtain the interference), as shown in Fig.~\ref{Chuu_fig17}. This is a standard Mach-Zehnder interferometer to the reference laser. Locking the phase difference of the two arms of the Mach-Zehnder interferometer with a feedback electronics stabilizes the phase of the two \keyword{SFWM} paths.  To avoid its interaction with the cold atoms, the reference beams are slightly displaced relative to the pump-coupling beams but pass through the same optical components.

By properly choosing the driving laser polarizations and the phase between the two \keyword{SFWM} paths, all four polarization-entangled Bell states can be realized for subnatural-linewidth \keyword{biphoton}s \cite{YanPRL2014}.

\section{\label{subsec:Applications}Applications}

\begin{figure*}
\includegraphics[width=\linewidth]{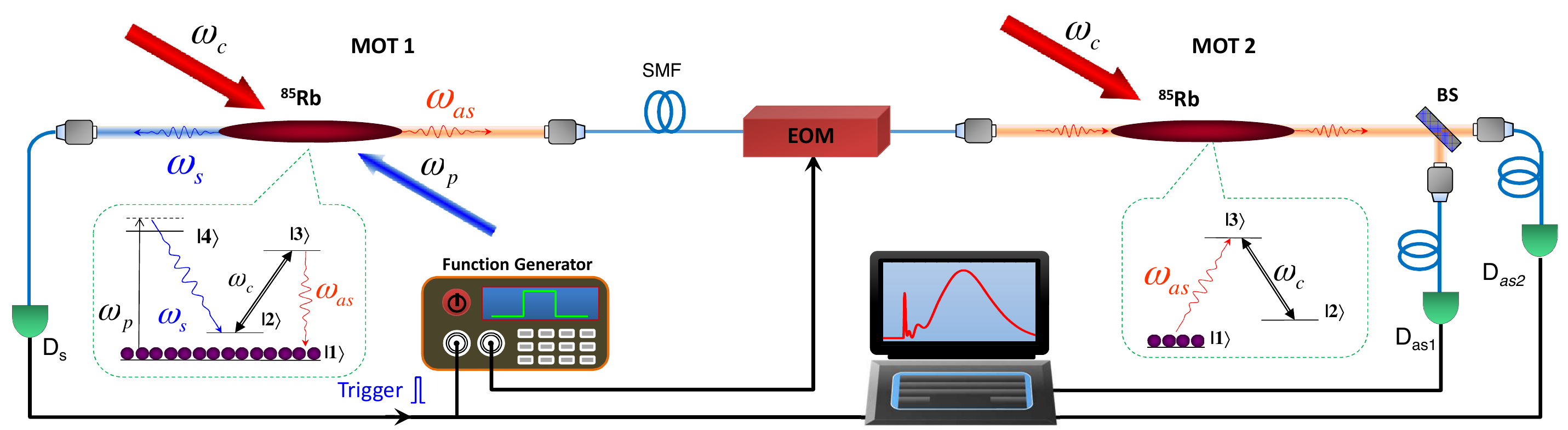}
\centering
\caption{\label{Chuu_fig18} Schematics of experimental setup for studying photon-atom quantum interaction. Narrow-band paired Stokes and anti-Stokes photons are produced from a cold atomic ensemble in MOT1. The anti-Stokes photons pass through an EOM driven by a function generator triggered by the detection of Stokes photons at D$_s$. We then send the heralded anti-Stokes photons with amplitude modulation to the second cold $^{85}$Rb atomic ensemble at MOT2. The figure is taken from Ref. \cite{SinglePhotonPrecursor}.}
\end{figure*}

In the past, experimental manipulation of a single photon interacting with atoms have been mostly focused in the frequency domain. Now with the narrowband \keyword{biphoton} generation technique described in this chapter, it is possible to control the quantum interaction between a single photon and atoms in time domain. In this section, we review the applications of heralded single photons whose waveforms are shaped by an EOM. A general schematics is illustrated in Fig.~\ref{Chuu_fig18}. We work with two-MOT setup. Narrowband Stokes and anti-Stokes \keyword{biphoton}s are spontaneously produced from cold atoms in the first MOT (MOT1). After detection of a Stokes photon, its heralded anti-Stokes photon, after shaping by an EOM, is directed to the cold atoms in the second MOT (MOT2). The atoms in MOT2 can be either a three-level \keyword{EIT} system with a coupling laser or a two-level system without the coupling laser. In the following we show three examples on how the photon-atom interaction in MOT2 can be precisely controlled by manipulating the temporal waveform of the heralded single photons.\\

\noindent\textbf{1) Optical Precursor of a Single Photon}\\

Now it has been well accepted that both phase velocity and group velocity of light in a dispersive medium can exceed $c$, the speed of light in vacuum \cite{Fastlight1, Fastlight2}. Then, what is the information velocity of light? Could it be faster than $c$ and violate Einstein's causality in the special relativity? Answering this question motivated early study of optical \keyword{precursor}s by Sommerfeld and Brillouin in 1914 \cite{Sommerfeld1914, Brillouin1914, Brillouin}. They showed theoretically that the front of a step-modulated optical pulse (as a carrier of 1 bit information) propagating in a dispersive medium always travels at $c$. This front, in the form of a transient wave now known as the Sommerfeld-Brillouin \keyword{precursor}, is then followed by the main pulse traveling at its group velocity.  Study of optical \keyword{precursor}s is of great interests not only for fundamental reasons since it is related to Einstein's causality, but also for applications because of its connection to the maximum speed of optical information transmission \cite{Oughstun, Aavikscoo1991, Choi2004, Heejeong2006, Wei2009, StackedPrecursor, ChenJOpt2010, DuPhysics2013}. Now it is clear that the \keyword{precursor} is the fastest part in the propagation of an optical pulse even in a superluminal medium \cite{Heejeong2006, Wei2009, ChenJOpt2010}. Classical \keyword{precursor}s are entirely based on macroscopic electromagnetic wave propagation.

A single photon is described by quantum mechanics. What is the speed of a single photon in a dispersive medium? In quantum mechanics, an observable physical quantity, such as the speed, usually takes many possible (discrete or continuous) eigenvalues. Some may argue that a single photon event may possibly violate Einstein's causality while the expectation value of its speed never exceeds $c$. Although The propagation effect of single photons through slow and fast light media has been studied previously \cite{LukinNature2005, ChiaoPRL1993, ChiaoPRA1995, AkopianNP2011}, all of these experiments focused on the group velocity picture. It remains a question whether there is a speed limit for a single photon in the quantum nonclassical world. The detection of single-photon \keyword{precursor} may shine light to this question as well as the understanding of quantum information transmission.

\begin{figure}[h]
\includegraphics[width=0.55\linewidth]{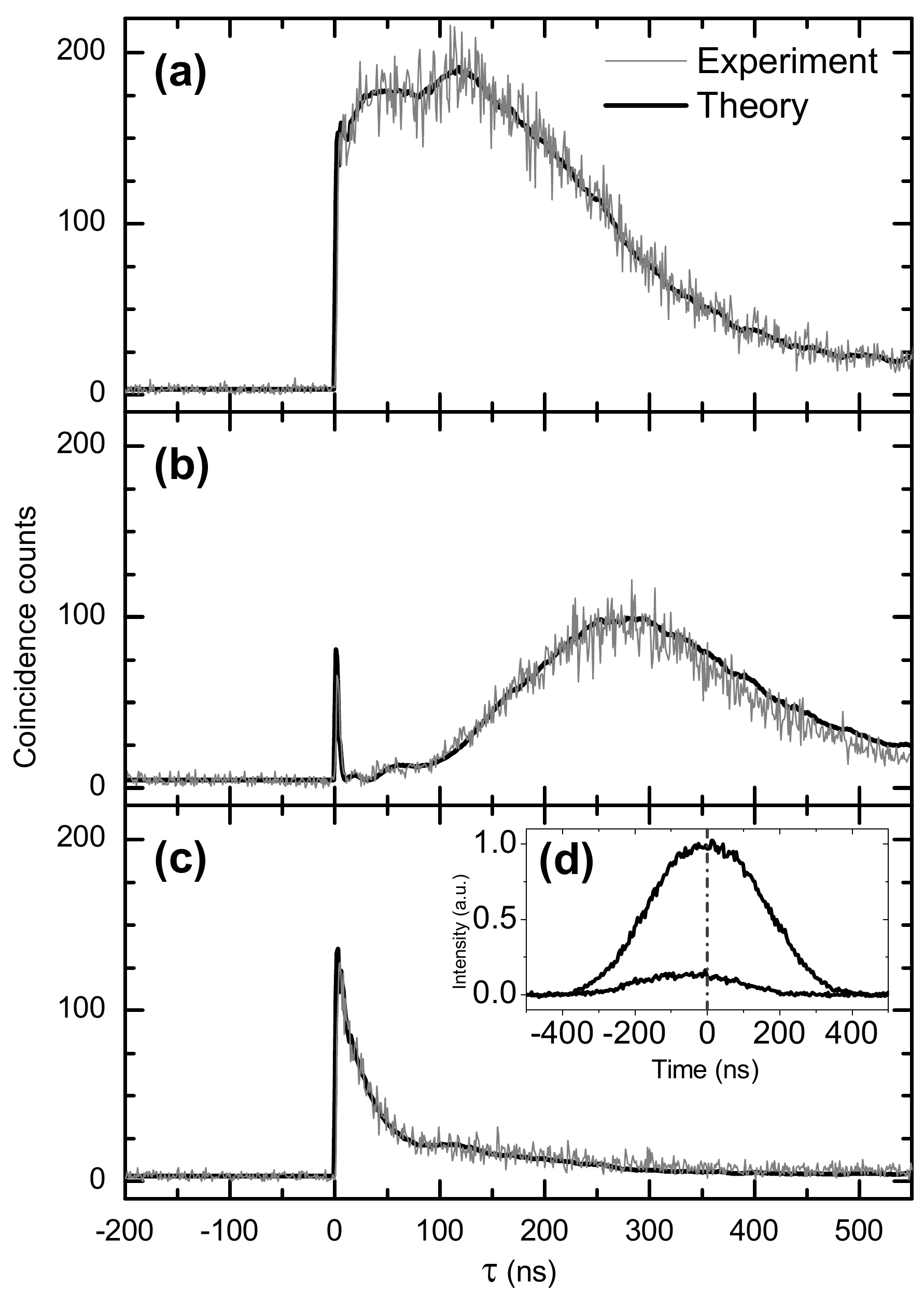}
\centering
\caption{\label{Chuu_fig19} Single-photon optical \keyword{precursor}s from a step amplitude modulation. (a) The heralded anti-Stokes photon waveform with a step modulation. (b) and (c) are two-photon coincidences after the anti-Stokes photons passing through the \keyword{EIT} system ($\Omega_{c2}=3.5\gamma_{13}$, OD=20) and two-level system ($\Omega_{c2}=0$, OD=2.5) in MOT2 respectively. Inset (d) shows a gaussian pulse propagation in the two-level system with a peak advancement of about 40 ns (the lower curve) compared to the reference pulse (the upper curve). Data are taken from Ref. \cite{SinglePhotonPrecursor}}
\end{figure}

Optical \keyword{precursor} of a single photon was first observed making use of narrowband heralded single photon with a step-shaped waveform \cite{SinglePhotonPrecursor}, with the experimental setup shown in  Fig.~\ref{Chuu_fig18}. The modulated heralded anti-Stokes photon waveform is shown in Fig.~\ref{Chuu_fig19}(a). After the anti-Stokes photon has passed through the \keyword{EIT} medium in MOT2 (OD=20), the \keyword{precursor} is clearly seen at the rising edge and separated from the delayed main waveform, shown in Fig.~\ref{Chuu_fig19}(b). To test single-photon causal propagation in a superluminal medium, we turn off the coupling laser in MOT2 to work in a two-level system, which has a negative \keyword{group delay}, as shown in Fig.~\ref{Chuu_fig19}(d). we observe a peak advance of at least 40 ns and with about 10\% transmission compared to the propagation through vacuum. However, as shown in Fig.~\ref{Chuu_fig19}(c), there is no observable advancement relative to the rising edge in the single photon waveform. That is, using the quantum mechanics interpretation, there is no observable probability for single photon traveling faster than $c$. This result indicates that the optical \keyword{precursor} is always the fastest part even in superluminal propagation and Einstein's causality holds.

The observation of optical \keyword{precursor} of heralded single photons \cite{SinglePhotonPrecursor} confirms the speed limit of a single photon in a dispersive medium, which is indeed the speed of light in vacuum. The information velocity of a single photon does not follow its group velocity. It also suggests that the causality holds for a single photon.\\

\noindent\textbf{2) Optimal Storage and Retrieval of Single-Photon Waveform}\\

The second application example of narrowband heralded single photons with arbitrary waveform shaping is for optical quantum memory. The efficiency of a photon-atom quantum interface strongly depends on the temporal shape of single photons \cite{LukinPRL2007}. Although storage of of weak coherent pulses have been demonstrated with high efficiencies up to 87\% \cite{Alexander, BCBucler2011}, obtaining such high efficiency with single photons remains a technical challenge due to the difficulty in having narrowband single photons with optimal temporal waveform at the operation frequency \cite{GisinNature2011, TittelNature2011, TittelPRL2012}.

\begin{figure}[htbp]
\centering\includegraphics[width=0.55\linewidth]{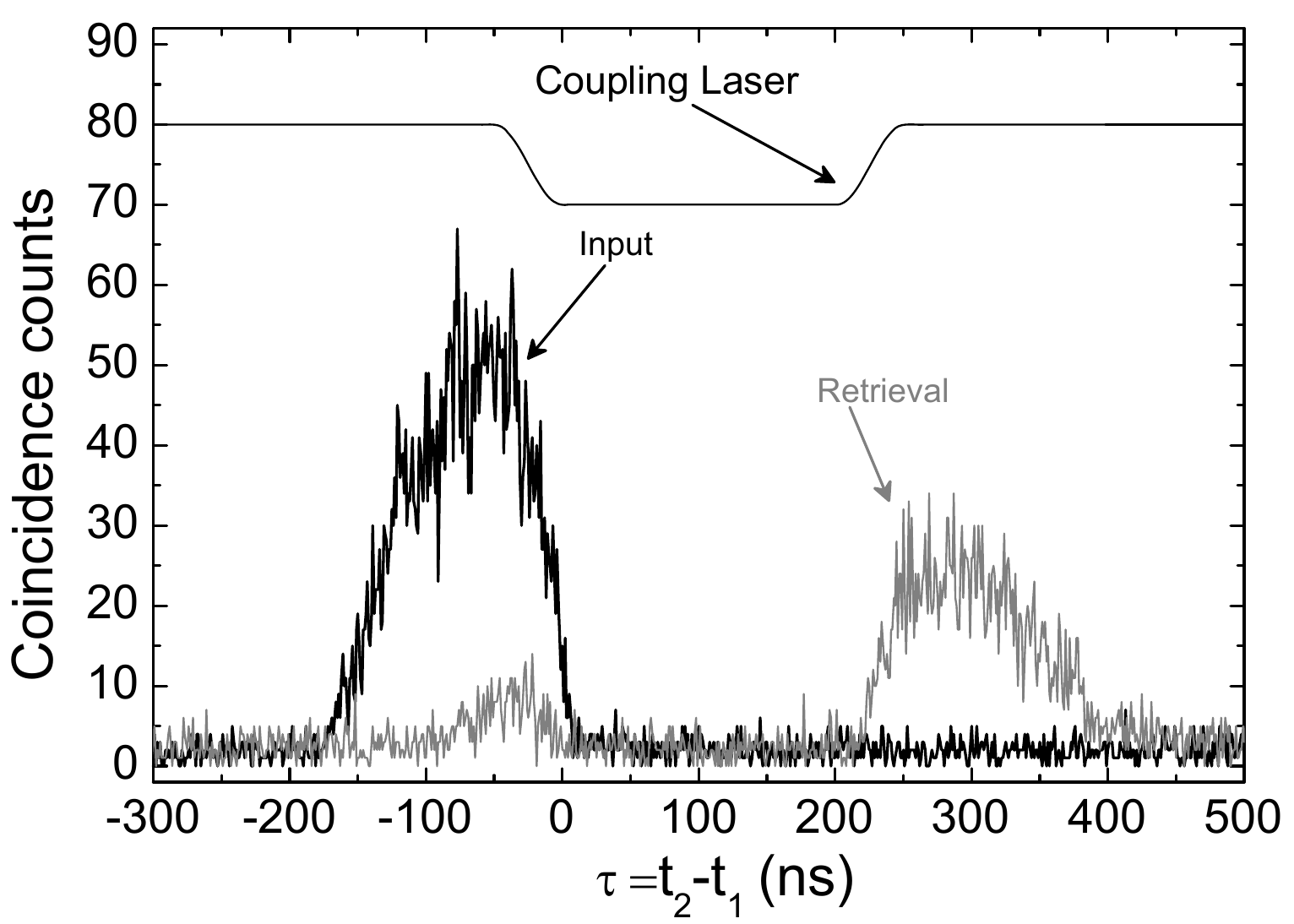}
\caption{\label{Chuu_fig20} Optimal storage and retrieval of single photons with a storage efficiency of (49$\pm$3)\%. (a) The optimal input (red curve) and output (retrieval, green curve) heralded single-photon waveforms. (b) The time-reversed retrieved photon waveform matches the input photon waveform after normalization. Data are taken from Ref. \cite{DuOptExpress2012}}
\end{figure}

Making use of heralded narrowband single photons with an optimal waveform shaped by EOM, Du \textit{et al} reported an experimental demonstration of efficient storage and retrieval of narrow-band single-photon waveforms using \keyword{EIT} in a cold atomic ensemble \cite{DuOptExpress2012}. With the ability to control both single-photon wave packets and the memory bandwidth, the storage efficiency is obtained up to (49$\pm$3)\%. The experimental result is shown in Fig. ~\ref{Chuu_fig20}. To our knowledge, it represents the highest storage efficiency for a single-photon waveform to date. Because an efficiency above 50\% is necessary to operate a memory for error correction protocols in one-way quantum computation \cite{VarnavaPRL2006}, this result brings the atomic quantum light-matter interface closer to practical quantum information applications.\\

\noindent\textbf{3) Coherent Control of Single-Photon Absorption and Reemission in a Two-Level Atomic Ensemble}\\

When an optical pulse propagates through a dispersive medium, the absorption and emission can coherently modify its spectral-temporal components and lead to many fundamental and important optical phenomena, such as attenuation, amplification, distortion, slow and fast light effects~\cite{EITHarris, slowlight, gaussianprop, twolevelamp, supsubreluminal}. For a single photon in the absorptive medium, the remission can only occur after the absorption, as required by the causality. Although this quantum time order has been observed as the antibunching effect in resonance fluorescence \cite{DuPRL2007, resfluor}, the processes in these experiments are chaotic. After a single photon enters a two-level atomic medium and gets absorbed, the spontaneously reemitted photon usually goes to $4\pi$ solid angle randomly.

It turns out a directed ``spontaneous'' emission excited by a single photon is possible if the timing of absorption is traceable \cite{ScullyPRL2006}. Moreover, this direct ``spontaneous'' emission can be manipulated in time domain by controlling the single-photon temporal waveform \cite{DuPRL2012}. For example, making use of the destructive interference between the emission (or scattering) and the incident photon wave packet, the probability of remitting the photon during the absorption can be completely suppressed when the incident photon has an exponential growth waveform with a time constant equal to the excite-state lifetime. The remission process only starts after the incident photon waveform is switched off and thus can be controlled on demand.

\begin{figure}
\includegraphics[width=0.55\linewidth]{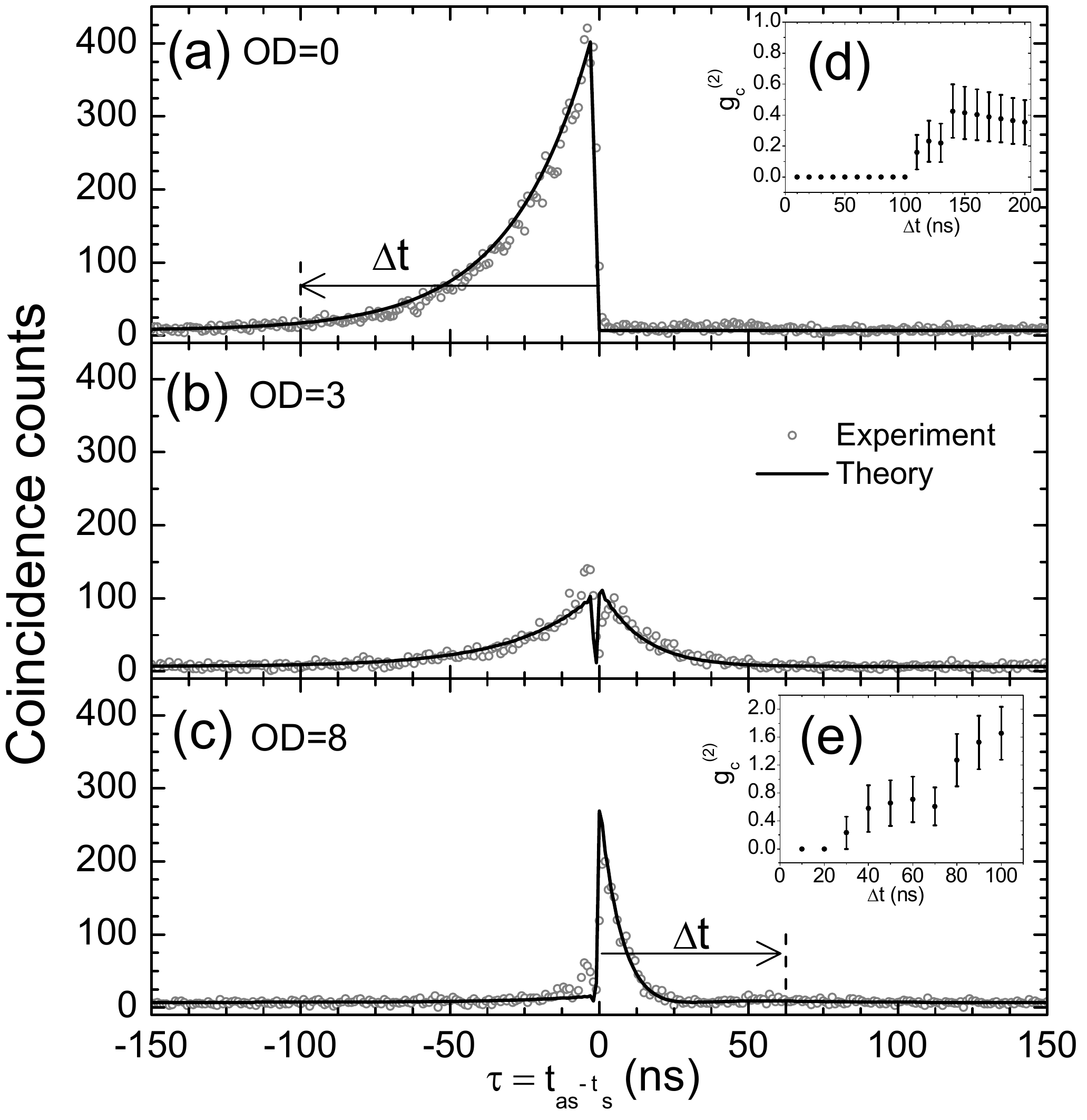}
\centering
\caption{\label{Chuu_fig21}Coherent control of single-photon absorption and reemission in a two-level laser-cooled atomic ensemble. Heralded anti-Stokes photon waveforms after passing through the two-level atomic ensemble at (a) OD=0 (vacuum), (b) OD=3, and (c) OD=8. (d) and (e) show the measured conditional autocorrelation $g^{(2)}_c$ for confirmation of the single-photon quantum nature. Data are taken from Ref. \cite{DuPRL2012}}
\end{figure}

We illustrate the process in Fig.~\ref{Chuu_fig18}, where we turn off the coupling laser in MOT2 and have the atoms there in a two-level system. With the EOM modulation, we produce single anti-Stokes photons with an exponential growth waveform with a time constant equal to $1/(2\gamma_{13})=$26.5 ns for the incident anti-Stokes photons, as shown in Fig.~\ref{Chuu_fig21}(a). Here $2\gamma=2\pi\times6$ MHz is the population decay rate in the excite state. At $\tau=t_2-t_1=0$, the waveform is switched off with a fall time of 3 ns. After passing through the atoms, the photons are coupled into a SMF to be detected by a SPCM. Fig.~\ref{Chuu_fig21}(b) shows the result at OD=3. At this modest OD, during the exponential growth period ($\tau<0$), the photon waveform is only partially absorbed. After the incident photon is switched off at $\tau$=0, this partially absorbed waveform is released (re-emitted) following a exponential decay curve which is determined by the lifetime [$1/(2\gamma)$] of the excited state. As we increase the OD, the incident photon gets absorbed more heavily. At OD=8, as shown in Fig.~\ref{Chuu_fig21}(c), the photon is completely absorbed and the probability in finding the remitted photon at $\tau < 0$ is nearly zero due to the destructive interference. As expected, at $\tau > 0$, the interference between the incident waveform and the emission disappears and we observe the remitted photon. Consequently, the absorption and reemission of the single photon is completely separated in time domain.

This technique can be used to efficiently excite a single quantum absorber in a cavity by a single photon \cite{JohnePRA2011, PinotsiPRL2008}. The result may find potential applications in the quantum networks which require efficient conversion between flying single-photon states and local atomic states \cite{QuantumInternet}.\\

\noindent\textbf{4) Single-Photon Differential-Phase-Shift Quantum Key Distribution}\\

\begin{figure}[htbp]
\centering\includegraphics[width=0.85\linewidth]{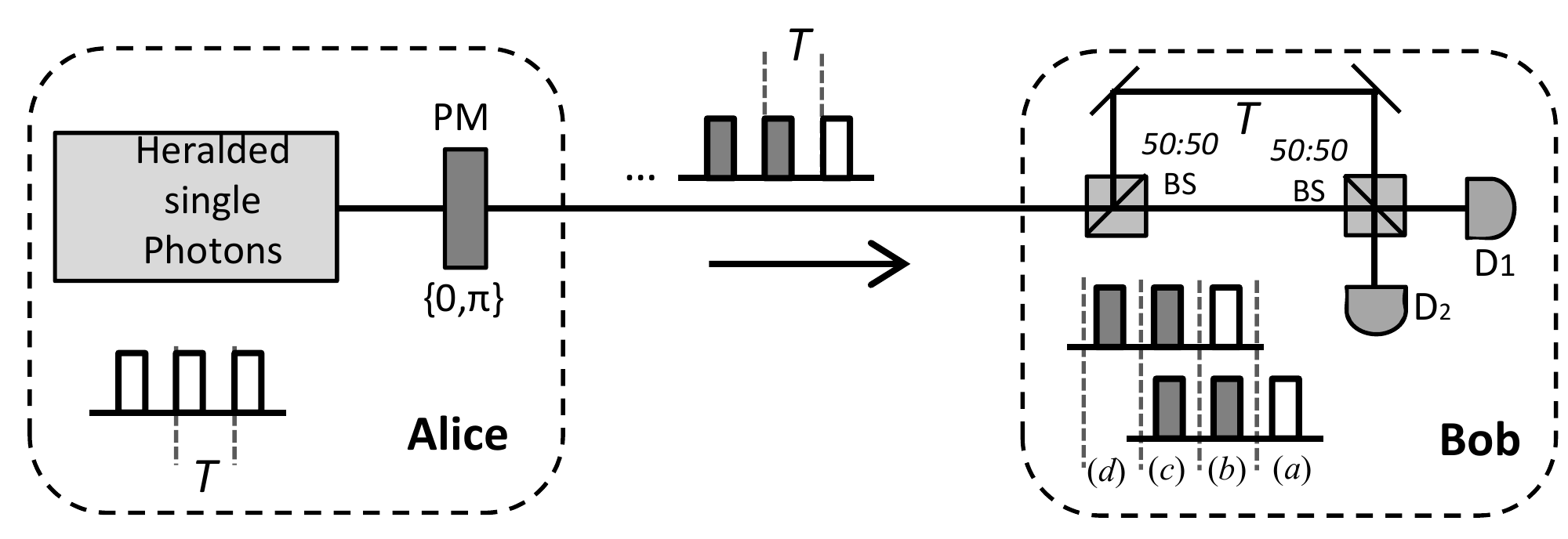}
\caption{\label{Chuu_fig22} Schematics of single-photon differential phase shift quantum key generation and distribution.}
\end{figure}

Due to the long coherence time of the heralded single photon, its wave packet can be encoded into many time bins with phase-amplitude modulation. It thus become an ideal information carrier for the differential phase shift quantum key distribution (DPS-QKD) \cite{InouePRL2002, Wen2009}. Traditionally, discrete polarization quantum states have been widely used due to their simplicity \cite{Bennett1984}. However, the fiber length of such a polarization-based QKD system is limited by the birefringence effect that causes the polarization fluctuation on the receiver. This limit can be overcome by the DPS-QKD scheme that is polarization independent. The DPS-QKD also shows tolerance to photon-number-splitting (PNS) attacks \cite{Wen2009,Waks2006}.

A simplified schematics of the first single-photon DPS-QKD system is illustrated in Fig.~\ref{Chuu_fig22} \cite{SinglePhotonDPS-QKD}. Alice divides the single photon into $N$ ($\geq 3$) time slots (with a period of $T$). The keys are encoded by the random relative phase shift between consecutive pulses in 0 or $\pi$. Bob detects the incoming photon using an unbalanced M-Z interferometer setup with a path time delay difference equal to $T$. Taking $N=3$ as an example, the detection at Bob's site yields four possible time-slot outputs, (a)-(d), as shown in Fig.~\ref{Chuu_fig22}. As Bob detect a photon, he records the time and which detector clicks. If the detector clicks at the (b) or (c) time slot, Bob tells Alice only the time slot information through a classical channel; otherwise, Bob discards the photon. Using the time-slot information and her phase encoding records, Alice knows which detector clicked at Bob's site. Defining the clicks at $D_1$ and $D_2$ as ``0" and ``1" respectively, Alice and Bob can obtain a confidential bit string as a sharing key. The photon sent from Alice to Bob can be written as one of the following four states: $ (|1_{1}0_{2}0_{3}\rangle\pm|0_{1}1_{2}0_{3}\rangle\pm|0_{1}0_{2}1_{3}\rangle)/\sqrt{3}$, where $1_{i=1,2,3}$ represents one photon at time slot \textit{i}. As nonorthogonal with each other, the four states cannot be perfectly identified by a single measurement, as shown by the noncloning theorem \cite{Gisin2002}, which guarantees the security of the scheme.

In the original DPS-QKD proposal \cite{InouePRL2002}, a single photon is split into $N$ paths with different lengths and then recombined with passive beam splitters. This brings unavoidable loss, which results in a low key creation efficiency [$\propto(N-1)/N^2$] when $N$ is large. Meanwhile, the phase stabilization between different paths becomes a serious problem $N$ increases. In the experimental demonstration of single-photon DPS-QKD, each heralded narrowband single photon is modulated into many time slots ($N$ up to 15) using a pair of phase and amplitude EOMs \cite{SinglePhotonDPS-QKD}. Without usind beam splitters, the entire key creation efficiency scales as $(N-1)/N$ and approaches 100\% at the limit of large $N$ \cite{SinglePhotonDPS-QKD, Yan2011}. The details of the experiment are described in Ref. \cite{SinglePhotonDPS-QKD}.

\section{\label{sec:Summary} Summary}

In summary, we have reviewed recent development in narrowband \keyword{biphoton} generation. For the \keyword{SFWM} in cold atomic medium, \keyword{EIT} effect is used not only for resonantly enhancing the $\chi^{(3)}$ nonlinearity by eliminating the resonance absorption but also for tuning the phase-matching bandwidth with its slow-light effect. For the monolithic resonant backward-wave and forward-wave SPDC with cluster effect and double-pass pumping, it is possible to realize a miniature ultrabright \keyword{biphoton} source. These narrowband \keyword{biphoton}s with long coherence time from tens nanoseconds to several microseconds can be used to produce heralded single photons with arbitrarily shaped temporal waveforms by phase-amplitude modulations. We also reviewed their applications in manipulating temporal quantum interactions between single photons and atoms, as well as in the quantum key distribution. Most recently, it was demonstrated that a single photon with time-reversed exponential growth waveform can be loaded into a single-sided Fabry P\'{e}rot cavity with near-unity efficiency \cite{DuPRL2014Cavity, SrivathsanPRL2014}. With the time-resolved quantum-state tomography \cite{TQST, MitchellPRL2014}, their further applications in quantum network and quantum information processing are to be explored.

\begin{acknowledgement}
C.-S. C. acknowledges support from the Taiwan Ministry of Science and Technology (NSC101-2112-M-007-001-MY3, MOST103-2112-M-007-015-MY3) and National Tsing Hua University (101N7014E1, 103N2014E1). S. Du acknowledges the support from the Hong Kong Research Grants Council (Project Nos. 601411, 601113, and HKU8/CRF/11G).
\end{acknowledgement}
\bibliographystyle{SpringerPhysMWM} 

\printindex
\end{document}